\documentclass{aa}  
\usepackage{graphicx}
\usepackage{txfonts}
\usepackage[figuresright]{rotating}
\begin{document}

\title{FADE, an instrument to measure the atmospheric coherence time}

\author{A. Tokovinin$^1$, A. Kellerer$^2$, V. Coud\'e Du Foresto$^3$}

   \offprints{A. Tokovinin}

\institute{Cerro Tololo Inter-American Observatory, Casilla~603, 
La Serena,  Chile \\
             \email{atokovinin@ctio.noao.edu}
\and
European Southern Observatory, 
Karl-Schwarzschild-Strasse, 2 D-85748 
Garching bei M\"unchen, Germany \\
              \email{aglae.kellerer@eso.org} 
\and
LESIA, Observatoire de Paris, section de Meudon, 5 place Jules Janssen, 92190 Meudon, France \\
              \email{vincent.foresto@obspm.fr} 
                           }

\authorrunning{Tokovinin, Kellerer, \& Coud\'e du Foresto}
\titlerunning{Fast Defocus Monitor}
%\titlerunning{FADE, an instrument to measure the coherence time}

   \date{Received ; accepted }

  \abstract
{}
{%aims
After proposing 
a new method of deriving the atmospheric time constant from the speed
of  focus variations (Kellerer \& Tokovinin
2007), we now  implement it with  the new instrument, FADE.
}
{%methods
FADE uses a 36-cm Celestron telescope that is modified to transform  
 stellar point images into a ring by increasing the central obstruction
 and combining defocus with spherical aberration.
Sequences of  images recorded  with a fast CCD detector  
are processed to determine the defocus and its variations in time from the ring radii. The
temporal structure function of the defocus is fitted with a model to
derive the atmospheric seeing and time constant. We investigated by numerical
simulation the data
reduction algorithm and instrumental biases. 
Bias caused by instrumental effects, such as optical aberrations,
detector noise, acquisition frequency, etc., is quantified.
The ring image must be well-focused, i.e. must have a sufficiently sharp radial profile, 
otherwise, scintillation seriously affects the results. An acquisition frequency of
700~Hz appears adequate. 
}
{%results
FADE was operated for 5 nights at the Cerro Tololo observatory 
in parallel with the regular site monitor. 
Reasonable agreement between the results from the two instruments has been obtained.
}
{}


   \maketitle

\section{Introduction}

The site- and time-dependent performance of telescopes, and especially
of  interferometers,  can  be  characterized by  the  parameters  {\it
seeing\/},   $\varepsilon_0$  (or,   equivalently,   the  {\it   Fried
parameter\/}  $r_0  =  0.98  \lambda/ \varepsilon_0$),  and  the  {\it
coherence  time\/}, $\tau_0$,  that determines  the  required reaction
speed of adaptive-optics  (Roddier \cite{Roddier81}).  The variability
of these parameters makes monitoring instruments essential.  Seeing is
usually  measured with  the  Differential Image  Motion Monitor,  DIMM
(Sarazin \&  Roddier \cite{DIMM}).  However, a  practical method
of measuring  $\tau_0$ is still  lacking.  At present this  parameter is
inferred from the vertical profiles of wind speed and turbulence, from
the temporal analysis of  image motion, from scintillation, etc.  (cf.
the review in Kellerer  \& Tokovinin \cite{Fade}, hereafter KT07).  In
particular, a  Multi-Aperture Scintillation Sensor,  MASS (Kornilov et
al.  \cite{MASS1}) deduces the  coherence time from scintillation, but
this method  (Tokovinin \cite{MASS_tau})  is only approximate  and has
not yet been verified by comparison with other techniques.

\begin{figure*}[ht]
\begin{center}
\includegraphics[width=16cm]{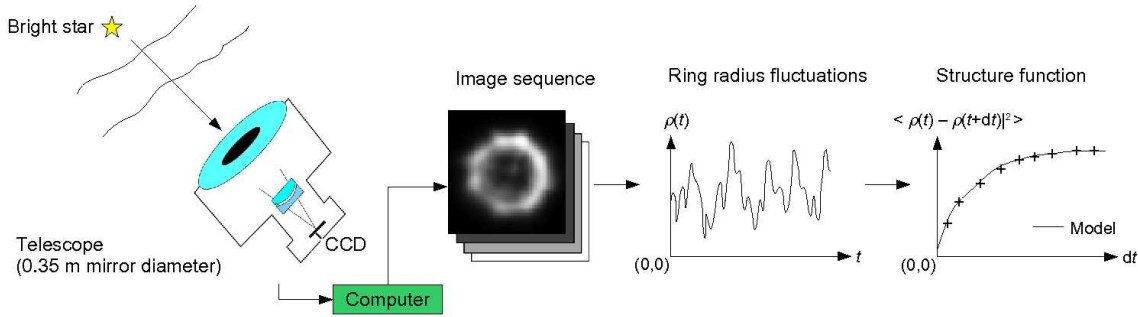} 
\caption{Overview of the FADE instrument and data analysis. }
\label{fig:setup} 
\end{center}
\end{figure*}

A new method for measuring the  coherence time with a small telescope has
recently  been  proposed in  KT07.   This  method,  termed FADE  (FAst
DEfocus),  is based  on  recording and  processing focus  fluctuations
produced  by the  atmospheric  turbulence in  a  small telescope.  The
variance   of  defocus  is   proportional  to   $(D/r_0)^{5/3}$  (Noll
\cite{Noll})  and gives  a measure  of the  seeing, where  $D$  is the
telescope diameter.   As shown in KT07,  the variance of  the speed of
defocus  is related  to  $\tau_0$  (this relation  is  given below  in
Sect.~\ref{sec:model}  where  the  derivation  of  $\tau_0$  from  a
sequence  of fast  defocus  measurement is  explained).  In  principle,
$\tau_0$  can be  obtained from  the temporal  analysis of  almost any
quantity  affected  by turbulence.   However,  tilts,  the easiest  to
measure,  are  typically  corrupted  by telescope  shake  and  guiding
errors, hence are not suitable.  The DIMM instrument is immune to
the wind shake, but it  is intrinsically asymmetric.  An early attempt
to  extract $\tau_0$ from  the DIMM  signal by  Lopez (\cite{Lopez92})
revealed  the complexity  of this  approach and  did not  result  in a
practical instrument.   If we discard  tilts, the next  second largest
and slowest  atmospheric terms  are defocus and  astigmatism.  Defocus
has  angular symmetry  and the  rate of  its variation  is  related to
$\tau_0$.  Attractive features of the FADE method thus are:
\begin{itemize}
\item
direct measurement of $\tau_0$ and $\varepsilon_0$,

\item
use of the whole telescope aperture,

\item
immunity to tilts,

\item
small telescope size.

\end{itemize}
FADE can be useful for site testing and monitoring, but its
 feasibility has so far been demonstrated only by numerical simulation.
Here we present an instrument implementing the new method.

The instrumental set-up  is described in Sect.\,\ref{Configuration}.
Section\,\ref{DataAnalysis}  outlines  the  data  analysis  algorithm,
while    various    instrumental     effects    are    evaluated    in
Sect.~\ref{Sec:Sim}      by      numerical      simulation.       In
Sect.\,\ref{statistics} the seeing  and coherence time measured with
FADE are checked for consistency and are compared to simultaneous data
from  the  DIMM   and  MASS  instruments.   Section\,\ref{Conclusions}
contains  conclusions and  an outline  of further  work.  Mathematical
derivations are given in Appendices~\ref{sec:radius} and \ref{sec:SF}.

\section{The instrument}\label{Configuration}

\subsection{Operational principle}\label{sec:principle}

The temporal structure function (SF) of atmospherically-induced focus
variations $D_4(t)$ is related to the average wind speed in the
atmosphere $\overline V$. The initial, quadratic part of SF is 
\begin{eqnarray}
D_4(t) \approx C t^2 \int_0^{+ \infty} {\rm d} h \;
C_n^2(h) \; V^2(h) =  C t^2 \overline{V}_2^2 \int_0^{+ \infty} {\rm d} h \;
C_n^2(h) ,
\label{eq:D4square}
\end{eqnarray}
where the  proportionality coefficient $C$ depends  on the wavelength,
telescope diameter  and central obstruction  (see KT07 and  Appendix B
for the  derivation), $C_n^2(h)$ and $V(h)$ are  vertical profiles
of the refractive-index  structure constant and wind  speed.  By measuring
SF, we  can estimate the integral in  (\ref{eq:D4square}) that is
similar  to  the  integral  entering   the  definition  of  $\tau_0$
(Roddier \cite{Roddier81}),
\begin{eqnarray}
\tau_0 = 0.314 r_0/\overline{V}_{5/3} =
\left( 118 \lambda^{-2}
 \int_0^{+ \infty} {\rm d} h \; C_n^2(h) \; V^{5/3}(h) \right) ^{-3/5} .
\label{eq:taudef}
\end{eqnarray}
The exponent of the wind speed is slightly  different; however,  we  show below
that fitting the measured  SF to a theoretical model leads to
a good estimation of $\tau_0$.

The defocus aberration can be measured with a wave-front sensor of
any  type or  can  be simply  inferred  from the  size  of a  slightly
defocused   long-exposure  stellar   image  (Tokovinin   \&  Heathcote
\cite{Donut}).   For FADE,  a  simple, fast,  and  accurate method  is
required.  We chose  to introduce a conic aberration  into the beam in
order to form a ring-like image.  A small defocus slightly changes the
radius  of the ring.   Ring-like images,  ``donuts'', are  obtained by
defocusing  a  telescope with  a  central obstruction.   However,
unlike  a donut, the  ring is  fairly sharp  in the  radial direction,
which  means that  the determination  of  the ring  radius is  largely
insensitive to intensity fluctuations (scintillation) at the telescope
pupil. 

 There is  an inherent similarity between FADE  and DIMM.  In a
DIMM, two peripheral beams are  selected and are deviated by prisms to
form an  image of two  spots.  In FADE,  the prisms are replaced  by a
conic aberration and the  whole annular aperture is used to form
a ring-like image  instead of two spots.

The ring  images are recorded by a  fast CCD detector and  stored on a
computer disk  (Fig.~\ref{fig:setup}). They are  processed offline to
determine a temporal sequence of ring radii, $\rho(t)$,  related to
the  defocus.   In  order  to  estimate  the  atmospheric  parameters
$\varepsilon_0$ and  $\tau_0$, the temporal structure  function of the
defocus variations is computed (Eq.~\ref{eq:D4def}) and fitted to a model.

Atmospheric defocus fluctuations are fast: their temporal correlations
decrease  with  a half-width  0.3  times  the  {\it aperture  crossing
time\/}  $t_{\rm cross}  = D/V$,  i.e.  with  2.2\,ms for  a telescope
diameter $D=0.36$\,m  and wind speed  $V =50$\,m/s (cf.   Appendix B).
To capture the focus  variations of interest, an acquisition frequency
$\nu \ge  500$\,Hz (exposure time $<2$\,ms)  is required, which
is attainable with today's fast CCD detectors.

\subsection{Hardware\/}
\begin{table}[ht]
\center
\caption{Components of the FADE instrument}
\label{tab:hard}
\medskip
\begin{tabular}{l  l }
\hline
\hline
Component & Description \\
\hline
Telescope & {\it Celestron\/} C14, $D=0.356$\,m, $F=3.910$\,m \\
Central obstruction & Circular mask of 150\,mm diameter \\
Aberrator & PCX lenses ({\it Linos\/} 312321 \& 314321), \\
 & $d_L=25$\,mm, $f_L=\pm50$\,mm\\
Detector & {\it Prosilica\/} GE\,680, $640\times 480$, \\
         & pixel 7.4\,$\mu$m ($0.39''$) \\
Interface & Gigabit Ethernet IEEE 802.3 1000baseT \\
Computer \& OS & {\it Dell\/} D410, {\it Windows\/} XP \\
\noalign{\medskip}
\hline
\hline
\end{tabular}
\end{table}

We  assembled the  FADE  prototype from  readily available  commercial
components (Table\ref{tab:hard}).  A 36-cm telescope was selected.  In
a smaller telescope, the  focus variations are smaller, $\langle a_4^2
\rangle \propto D^{5/3}$, and faster, $t_{\rm cross} \propto D$, hence
more difficult  to measure.  Use  of a fast  CCD -- GE\,680  from {\it
Prosilica\/}  -- is critical  for the  instrument, because  it permits
continuous acquisition  with an image frequency 740~Hz  when a 100x100
subsection of  the full frame  (region-of-interest, ROI) is  read out.
The signal  is digitized  in 12~bits.  With  the lowest  internal gain
setting, 0~dB,  the conversion factor  2.86\,ADU per electron  and the
readout  noise  38\,ADU~=~13.4\,e were  measured.   According to  the
specifications, the  maximum quantum efficiency (QE)  is 0.5 electrons
per  photon  at  wavelength  $\lambda=0.50\;\mu$m  with  a  full-width
half-maximum spectral  response of roughly $\Delta\lambda=0.25\;\mu$m.
Indeed,  the measured  fluxes  from stars  correspond  to the  overall
system QE of 0.35--0.40, including atmospheric and optical losses.

\subsection{Optics\/}

\begin{figure}[ht]
\begin{center}
\centerline{\includegraphics[width=8.5cm]{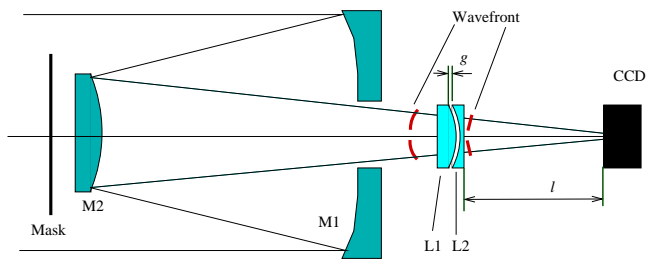}}
\caption{ Optical layout of FADE (not to scale). The stellar light
  comes from left to right. After reflection from the telescope
  primary and secondary mirrors M1 and M2, the converging beam passes
  through the ``aberrator'' consisting of the lenses L1 and L2 and is
  detected by the CCD.  The wave-front cross-sections before and after
  the aberrator are illustrated by  thick line segments. 
}
\label{fig:scheme} 
\end{center}
\end{figure}
To create annular  images, a conic aberration must  be introduced into
the  beam.   Conic lenses,  {\it  axicons},  have  wide technical  and
research applications and are commercially available.  
 In the case of FADE, we can successfully approximate a conic wave-front
by a combination of  a quadratic (defocus) and higher-order (spherical
aberration) terms.   

An  aperture with a  relative central  obstruction $\epsilon$  gives a
ring-like defocused image with the  average angular radius $\rho = a_4
\sqrt{48}  /[D  (  1  -   \epsilon)]$,  where  $a_4$  is  the  defocus
coefficient  in linear  units  (throughout this  article, the  Zernike
aberration coefficients  are given in the  Noll \cite{Noll} notation).
Such a ring  can be  sharpened by adding spherical
aberration.   Elementary analysis  shows that  for a  relative central
obstruction $\epsilon \sim 0.4$, the deviation from a conic surface is
minimized  when   $a_{11}  =  -0.1\;a_4$.   The  rms   error  of  this
approximation is $0.01  a_4$.  In order to get a  ring radius of $5''$
with  a 35-cm  telescope,  we need  the  defocus amplitude  $a_4  \approx
600$\,nm; hence,  the difference of the approximated  wave-front from a
perfect cone will  be 6\,nm, or $\lambda/80$ at  $\lambda = 500$\,nm.

  To attain the desired aberration,  we used an assembly of two simple
plano-spherical lenses with equal  but opposite curvature radii, which
can  be seen as  a plane-parallel  plate containing  a meniscus-shaped
void  (Fig.~\ref{fig:scheme}).   The  thickness  of  the  meniscus  is
adjusted by  changing the  gap $g$ between  the lenses.   The positive
lens is closer  to the primary mirror, so  that the meniscus curvature
opposes the  curvature of  the wavefront.  We  used lenses  with focal
lengths $f_L=\pm50$\,mm  and a gap $g=0.7$\,mm.  When  this element is
placed  at distance  $l=93.5$\,mm in  front  of the  detector and  the
telescope is suitably refocused, a  ring image of radius $\rho \approx
5''$ is  formed.  Optical  modeling in {\it  Zemax\/} shows  that this
``aberrator"  is reasonably achromatic.   The spherical  aberration is
proportional to $g  \, l^6$, therefore it can be  adjusted over a wide
range. To block the inner part of the wavefront  where it deviates
from the  cone, a  central obstruction of  150\,mm diameter,  i.e.  a
relative  diameter  $\epsilon=0.42$,   was  placed  at  the  telescope
entrance.    Obviously, there  are  many  other possible  optical
arrangements to obtain wave-fronts with spherical aberration.

\begin{figure*}[ht]
\begin{center}
\centerline{\includegraphics[width=12cm]{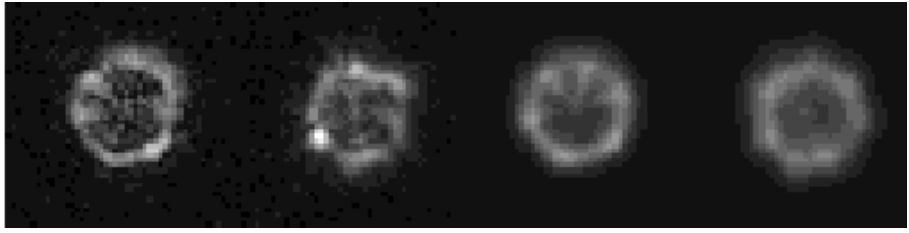}}
\caption{From left to right: Simulated ring image -- Image of Sirius  
-- Average of 1024 simulated images -- Average of 1024 Sirius images.
The sequence of Sirius images was recorded on Nov.\,2 at 6:46\,UT.
The parameters for the data and simulations are given in Sect.\,\ref{Sec:rings}.
}
\label{fig:rings} 
\end{center}
\end{figure*}

 The   average    ring   image    in   the   real    FADE   instrument
(Fig.~\ref{fig:rings})  shows marked  aberrations other  than conical,
caused by the  defects of optical surfaces and  of alignment.  Similar
rings were  reproduced in our  simulations with a combination  of coma
and  higher-order aberrations  (cf.   Sect.~\ref{Sec:rings}). We  also
fitted  the  Zernike  aberrations  directly  using  the  donut  method
(Tokovinin  \&  Heathcote  \cite{Donut})   and  found  that  the  coma
coefficient  could  reach   $\sim$100\,nm  (1.2\,rad).   Furthermore,  the
defocus was  not always kept at  its optimum value as required for sharp
ring  images.    The  effect  of   such  aberrations  is   studied  in
Sect.~\ref{Sec:uncts} by simulation.

\subsection{Acquisition software\/}

Since the GE\,680  detector is relatively new, with  no readily available
software development kits as yet, we used the commercial software,
{\it  Streampix\/} from  {\it Prosilica}.   It provides  all necessary
functions for  detector control and  data storage in the  FITS format,
but the parameters need to  be set manually at each acquisition, which
requires constant  attention.  And they  are not logged into  the FITS
headers  or otherwise.   Thus,  {\it Streampix}  is  only a  temporary
solution.  We checked  that the image sequence is  acquired at regular
intervals,  without time  jitter. The  detector was  exposed  for this
purpose to a strictly periodic light signal at 10\,Hz, and a $100\times
100$  ROI  was  read at  400\,Hz.   The  power  spectrum of  the  flux
calculated  from these data  is a  narrow peak  at $(10.0\pm0.2)$\,Hz
without significant tails.

\subsection{Observations}

The FADE instrument  was  installed in the USNO  dome of the Cerro
Tololo  Inter-American  Observatory (CTIO)  in  Chile  for the  period
October 27 to  November 3, 2006.  The  instrument was elevated 
$\sim 4$\,m above the ground.   We pointed FADE at bright stars, {\it
Fomalhaut\/} ($\alpha$~PsA, A3V, $m_V=1.16$) in the evening, then {\it
Sirius\/}  ($\alpha$~CMa, A1V,  $m_V=-1.47$).   The  exposure time
ranged  from 1\,ms  to 1.9\,ms  for Fomalhaut  and was  $<0.5$\,ms for
Sirius  to avoid  saturation.   Figure~\ref{fig:rings} shows  typical
instantaneous  and average  images  of Sirius,  as  well as  simulated
images.   During our test  run, the  seeing was  not very  good at
roughly $1''$,  
and the turbulence  in the high atmosphere  was strong
and fast, as shown by the MASS data.

\section{Data analysis\/}
\label{DataAnalysis}

A correct algorithm of  data processing and interpretation is critical
for deriving the  atmospheric parameters $\varepsilon_0$ and $\tau_0$.
We  carefully   selected  the   most  robust  method   of  calculating
atmospheric  defocus  from the  ring-like  images  and used  numerical
simulation  to   study  the  influence  on  the   results  of  various
instrumental effects  and of optical  propagation (Sect.~\ref{Sec:Sim}
below).

\subsection{Estimating the ring radius}
\label{Sec:radius}

The center  of gravity of the  image $(x_c,y_c)$ is  calculated by the
usual formula
\begin{eqnarray}
x_c  = \sum_{l,k} x_{l,k}\;I_{l,k} /   \sum_{l,k} I_{l,k}  \hskip .2 in
{\rm and \/} \hskip .2 in   y_c  = \sum_{l,k} y_{l,k}\;I_{l,k} /   \sum_{l,k} I_{l,k} . 
\label{eq:centercoord}
\end{eqnarray}
The ring radius $\rho$ can then  be estimated in a similar way, as the
intensity-averaged distance from this center:
\begin{eqnarray}
\rho  = \sum_{l,k} r_{l,k}\;I_{l,k} /   \sum_{l,k} I_{l,k}  .
\label{eq:rho}
\end{eqnarray}
Here $ I_{l,k}$ is the  light intensity at pixel $(l,k)$, and $r_{l,k}$
is the distance of this pixel from the center.  There are various caveats
below the apparent simplicity of this procedure.

There is  no unambiguous way to  assign a center to  a real (distorted
and  noisy) ring  image.  A  simple center-of-gravity  is a  very rough
estimate  of  $(x_c,  y_c)$;  in  particular, it  is  affected  by  the
intensity fluctuations in the ring  due to scintillation. It is better
to compute $(x_c, y_c)$ with  clipped intensities: 0 below a threshold
and 1 above,  the threshold being set safely  above the background and
its fluctuations.   This initial estimate  can be  improved further by
minimizing the intensity-weighted mean distance of the pixels from the
ring, as described in Appendix  A.  However, small inaccuracies in the
center  determination do  not affect  the resulting  radius critically,
and in fact, we found the initial estimate to be adequate.

A second caveat concerns the choice  of the pixels used for the radius
estimate.  A considerable  fraction of pixels lie outside  the ring in
an empty area that only contributes noise.  To reduce the noise with a
minimal loss  of information,  we  restricted  the pixels  used in
(\ref{eq:rho})  to a  mask  of inner  radius $\overline{\rho}  -\Delta
\rho$   and  outer  radius   $\overline{\rho}  +\Delta   \rho$,  where
$\overline{\rho}$  is the average  ring radius.   We express  the mask
half-width  $\Delta \rho$ as  a fraction  $\delta$ of  the diffraction
half-width of the ring,
\begin{equation}
 \Delta \rho = \delta  \;\lambda/[0.5\;D\;(1-\epsilon)] .
 \end{equation}

\begin{figure}[ht]
\begin{center}
\centerline{\includegraphics[width=8.5cm]{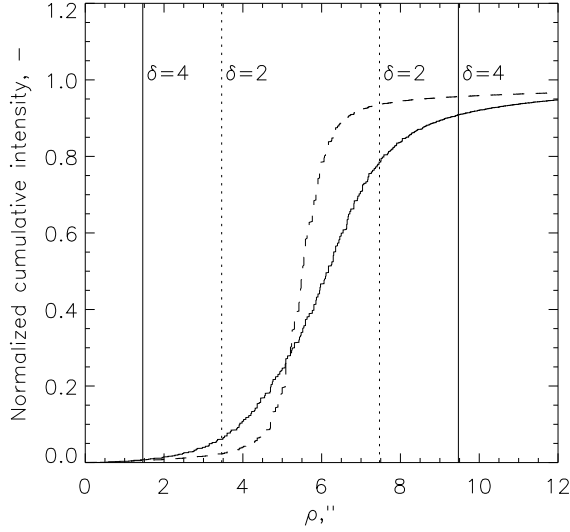}}
\caption{Total  intensity inside concentric circles of  radii  $\rho$
  for the average of 1024 centered images.  
  Full line: sequence of Fomalhaut images recorded on Nov.\,2. 
  Dashed line: simulated diffraction-limited ring images (see Table\,\ref{tab:simpar}).
    }
\label{fig:mask} 
\end{center}
\end{figure}

Figure\,\ref{fig:mask} shows that a mask with $\delta=2$ would be good
for an ideal, diffraction-limited ring. For a typical image sequence,
however, the ring is  widened by telescope aberrations and atmospheric
distortions, so we  set $\delta=4$, which covers the  actual ring image
with a sufficiently conservative, but still reasonable, margin.
 
The simulations  show that scintillation  and aberrations add  to the
fluctuations  of the  estimated radii  and  thus bias  the results  of
FADE.  To  reduce this  effect,  we  subdivide  the ring  into  eight
45$^\circ$  sectors   and  --  utilizing  the   same  center  estimate
$(x_c,y_c)$ -- apply Eq.\,\ref{eq:rho}  to each sector separately, and
then  average  the  result.   This  reduces the  effect  of  azimuthal
intensity variations.  An added advantage of the procedure is that the
relative variance  $s$ of the  total intensities in the  sectors $I_k$
with respect to their average  $\overline{I}_k$ serves as a measure of
the scintillation, hence of the turbulence height,
\begin{eqnarray}
s&=& \frac{1}{8} \sum_{k=1}^8 (I_k - \overline{I}_k )^2  \;/ \;
\overline{I}_k \;^2 .
\label{eq:sect}
\end{eqnarray}

The method of calculating the ring parameters $(x_c,y_c,\rho)$ is less
rigorous than  fitting a wave-front  model directly to the  image. The
big  advantage  of the  estimator  (\ref{eq:rho}),  however, is  its
simplicity.

\subsection{Noise and limiting stellar magnitude}

  The errors of the radius  estimates caused by photon and readout
noise are obtained by differentiating  Eq.\,(\ref{eq:rho}) and
using the independence  of the noise in each  pixel:
\begin{eqnarray} 
\sigma^2_{\rho, {\rm noise}} & = &   \left( \frac{\sigma_{\rm ron}}{N_{\rm ph}} \right) ^2 
\sum_{l,k} (r_{l,k} - \overline{\rho})^2 +
\frac{\delta^2_{\rho}}{N_{\rm ph}} ,  \label{eq:noise}\\
\delta^2_{\rho} &=& \sum_{l,k}  I_{l,k}\;(r_{l,k} - \overline{\rho} )^2 \; / \; N_{\rm ph} .	
\label{eq:width}
\end{eqnarray}
Here,   $\sigma_{\rm  ron}$   is  the   rms  detector   noise,  $N_{\rm
ph}=\sum_{l,k}  I_{l,k}$  the total  stellar flux  in  one exposure
(both in electrons), $\overline{\rho}$  the average ring radius, and
$r_{l,k}$ is the distance of  pixel $(l,k)$ from the center, expressed
either  in pixels  or  arc-seconds.  The  rms ring-width  $\delta_\rho$
quantifies  the ring  sharpness, which  turns  out to  be critical  for
getting unbiased  measurements with FADE  (see Sect.~\ref{Sec:uncts}).  The
summation is extended  only over pixels inside the  mask, as described
in  Sect.\,\ref{Sec:radius}.  We  recognize  a familiar  sum of  the
readout noise (first  term) and photon noise (second  term), where the
first term typically dominates.  Equation~(\ref{eq:noise}) does not account
for such additional noise  sources as scintillation, image distortion,
etc.

Formula  (\ref{eq:noise})  is   useful  for  predicting  the  limiting
magnitude of FADE. A star  of zero $V$-magnitude gives a flux $N_{\rm
ph}  \sim   6\cdot10^5$  photo-electrons  in  1\,ms   exposure  in  our
instrument.   The rms  noise  on the  radius  estimate with  plausible
parameters ($\rho = 5''$, $\sigma_{\rm  RON} = 13$, $\delta = 4$) is
then  about 2\,mas.  It  will increase  to 20\,mas  for a  star with
$m_V=2.5^m$  -- still much  less than  the atmospheric  signal. 
Despite very short exposures, FADE is not photon-starved.

\subsection{The response coefficient of FADE}

\begin{figure}[ht]
\begin{center}
\centerline{\includegraphics[width=8.5cm]{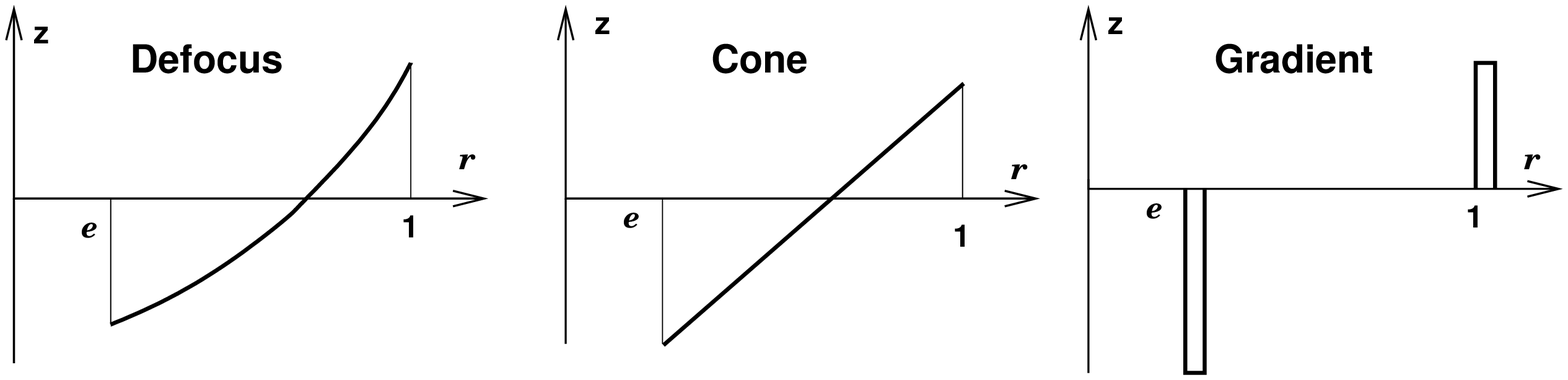} }
\centerline{
\begin{tabular}{| l  l | }
\hline
Defocus & $z(r) =  \sqrt{12}\left( r^2 - \frac{1 + \epsilon^2}{2} \right)/(1 -
\epsilon^2)$ \\
Cone & $ z(r) =  \left[ r - \frac{ 2(1 - \epsilon^3)}{3(1 -
  \epsilon^2)} \right] \left[ \frac{1 + \epsilon^2}{2} - \frac{ 4 ( 1
    - \epsilon^3)^2}{9(1 - \epsilon^2)^2} \right] ^{-1/2}$ \\
Gradient & $z(r) =  \left[ \frac{\delta(r-1)}{2 \pi r} - \frac{\delta(r -
    \epsilon)}{2 \pi \epsilon r}
   \right] $ \\ 
\hline
\end{tabular}
}
\caption{ Response  functions $z(r)$  on annular aperture  for Zernike
defocus, conic aberration, and  average radial gradient. The first two
functions  are  normalized  in  the  Noll  (\cite{Noll})  sense.   The
coefficients  $a_4$,  $a_c$,  and  $a_g$ are  calculated  as  integrals
(\ref{eq:a4}).   Here $\delta$ is the Dirac's delta function.  }
\label{fig:defocus} 
\end{center}
\end{figure}

The relation  between the ring  radius fluctuations $\Delta  \rho$ and
the  atmospheric defocus  (Zernike coefficient  $a_4$)  is intuitively
clear.  But  what is  the {\it exact}  coefficient $A$ in  the formula
$\Delta \rho  = A a_4$? Recall  that the atmsopheric  defocus $a_4$ is
related to the phase distortion $\varphi({\bf  r})$ as  

\begin{equation} 
a_4 = \int  {\rm d}^2 {\bf r} \; z_4({\bf r}) \; \varphi ({\bf r}) ,
\label{eq:a4}
\end{equation}
where ${\bf  r}$ is the normalized coordinate  vector on the pupil
and $z_4({\bf  r})$   the orto-normal  Zernike defocus  given by
Noll    (\cite{Noll})    for   the    circular    aperture   and    in
Fig.~\ref{fig:defocus} for the annular aperture.

A reaction  of our simple  radius estimator (\ref{eq:rho}) to  a small
perturbation  of phase  and amplitude  at the  telescope pupil  can be
determined  analytically (cf.   Perrin  et al.   \cite{Perrin} for  an
example of  similar analytics).  It turns  out that the  response to a
phase perturbation in  the pupil plane is not  exactly proportional to
the  Zernike  defocus.   Moreover,  it  depends on  the  adopted  mask
half-width $\delta$.   For $\delta \sim  1$, the response  resembles a
cone,   so that  FADE   measures  something   similar  to   a  conic
aberration. On  the other hand, for  $\delta \ge 2$  the computed ring
radius is  related to the  average radial gradient of  the wave-front,
and  therefore FADE measures  the difference  $a_g$ between  the phase
averaged on  the outer and inner  edges of its  annular aperture.  Its
response  is   further  modified  when   the  ring  is   distorted  by
aberrations.  In this  case, the radius estimate is  sensitive to both
amplitude  and  phase  fluctuations.   Although we  developed  a  full
analytical treatment of this problem,  it is omitted here for the sake
of simplicity.

The three quantities -- Zernike defocus $a_4$, conic aberration $a_c$,
and average  phase gradient  $a_g$ -- are  similar, especially  on the
annular aperture (Fig.~\ref{fig:defocus}). FADE measures  something
else, but its response is  most closely approximated by $a_g$ when the
ring radius is calculated with a large mask width $\delta$.  Let $a_g$
be the average phase difference  between the outer and inner borders of
the aperture, the  corresponding change of the angular  ring radius is
then
\begin{equation} 
\Delta \rho = \frac{a_g \;  \lambda}{ \pi D (1 - \epsilon)} .
\label{eq:Drho}
\end{equation}
The Zernike  defocus on the  annular aperture is proportional  to $a_4
\sqrt{12}r^2  / (1 -  \epsilon^2) $,  where $r$  is normalized  by the
pupil  radius.    Hence,  $a_g  =   a_4  \times  \sqrt{12}$,   and  the
proportionality coefficient $A$ follows from Eq.\,(\ref{eq:Drho}),
\begin{equation} 
\Delta \rho = A \; a_4 = a_4 
\frac{\lambda}{\pi D} \;   
  \frac{\sqrt{12}} {1 - \epsilon} .
\label{eq:A}
\end{equation}
The atmospheric  variance of  the defocus $a_4$  or gradient  $a_g$ on
an annular aperture  can be computed,  as done by Noll  (\cite{Noll}) for
a filled aperture.  Alternatively, the variance of the  ring radius may
be directly written as
\begin{equation} 
\sigma_\rho^2 = C_\rho (\lambda/D)^2 (D/r_0)^{5/3}
\label{eq:Crho}
\end{equation}
analogous to  similar  formulae for the  gradient or  Zernike tilt.
Our   numerical   calculation   for  the   average-gradient   response
(see Eq.\,\ref{eq:Drho}) leads to an approximation 
valid for $\epsilon < 0.6$ with an accuracy of $\pm 7\,10^{-5}$:
\begin{equation}
C_\rho \approx  0.03288 + 0.0503 \epsilon - 0.05638 \epsilon^2 + 0.04056
\epsilon^3.
\label{eq:Crho1}
\end{equation}

We  studied  the  response  of  FADE  by  analytical  calculation  and
numerical simulations and found that the exact coefficient $C_\rho$ in
Eq.\,\ref{eq:Crho} depends  on all parameters of the  instrument and data
processing.  A choice of $\delta \ge 2.5$ ensures a relative stability
of the response with respect to small aberration, propagation, etc.  A
small  correction to  the  ``ideal'' response  coefficient is  finally
determined  by simulation (Sect.~\ref{sec:Acoef})  and applied  to the
real data.

The   lack  of   a  unique,   well-established   coefficient  relating
measurements to atmospheric parameters may appear disturbing. However,
a similar analysis  applied to the classical DIMM  instrument leads to
the  conclusion that  its response,  too,  depends on  the details  of
centroid calculation and, furthermore,  is modified by propagation and
optical aberrations. In this respect, FADE and DIMM are not different.

\begin{table*}[ht]
\center
\caption{Simulation parameters.
}
\label{tab:simpar}
\medskip
\begin{tabular}{l ccc ccc c c c cccc  }
\hline
\hline
& $N$ & $\nu$      & d$t$   & $m_V$ &  $\sigma_{\rm ron\/}$ & 
$\rho$ & $a_4$  & $a_7$ &$a_{11}$ &$a_{27}$ & $h$ & $\varepsilon_0$ & $\overline{V}$ \\ 
       &     &  Hz        & ms &         &    el.                &
 $''$  & rad    & rad   &  rad   & rad  &  km & $''$   & m/s  \\
\hline
Fig.\,\ref{fig:rings}  & 1024 & 700 & 0.15 & -1.5 & 17 & 3.8 & 0& 0.7& -0.75& 0.3 & 13 & 1.05 & 17 \\
Fig.\,\ref{fig:mask} & 1024 & 700 & 0.15 & -1.5 & 17 & 5.5 & 0 &0 &0 &0  & 5 & 1.00 & 17 \\
Eq.\,\ref{eq:corr} & 1024 & 700 & 0.15 & -1.5 &  17 & 3.8 & 0&  0&  0&  0 & 10 & var. & 35 \\

Fig.\,\ref{fig:spherical} top & 1024 & 700 & 0.15 & -1.5 &  17 & 0 & 12&  0.7&  var. &  0.3 & 0 & var. & 35 \\ 
Fig.\,\ref{fig:spherical} bottom & 1024 & 700 & 0.15 & -1.5 &  17 & 0 & 12&  0.7&  var. &  0.3 & 5 & var. & 35 \\ 

Fig.\,\ref{fig:uncertainties} left & 1024 & 700 & 1.4 & var. &  17 & 4 & 0&  0&  0&  0 & 5 & var. & 35 \\ 
Fig.\,\ref{fig:uncertainties} middle & 1024 & 700 & 0.15 & -1.5 &  17 & var. & 0&  0&  0&  0 & 5 & var. & 35 \\ 
Fig.\,\ref{fig:uncertainties} right & 1024 & 700 & 0.15 & -1.5 &  17 & 4 & 0&  var. &  0&  0 & 5 & var. & 35 \\ 

\noalign{\medskip}
\hline
\hline
\end{tabular}
\end{table*}

\subsection{Derivation of the seeing and coherence time}
\label{sec:model}
\begin{figure}[ht]
\begin{center}
\centerline{\includegraphics[width=8.5cm]{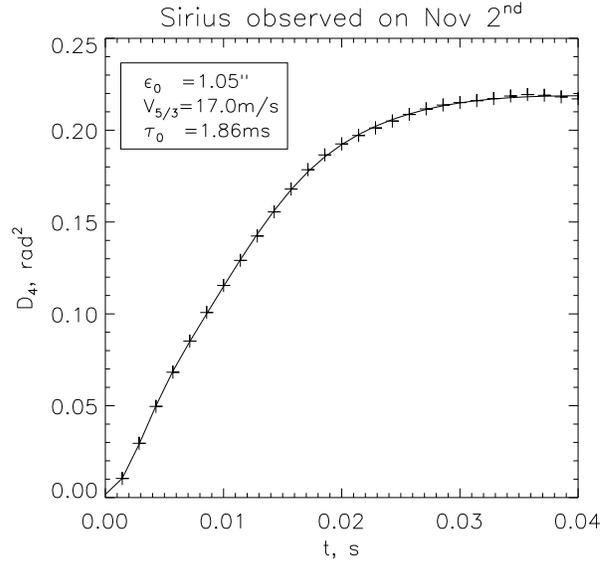} }
\caption{Structure function of focus variations measured at 700\,Hz
  (crosses) fitted with a model of three turbulent layers (line).
}
\label{fig:SF} 
\end{center}
\end{figure}
We convert the measured ring radius into defocus using coefficient
$A$ (see Eq.~\ref{eq:A})  and calculate the  temporal structure function
of defocus $D_4(t)$, 
\begin{equation}
D_4(t) = \langle [a_4(t' + t) - a_4(t')]^2 \rangle .
\label{eq:D4def}
\end{equation}
A typical SF is plotted in Fig.~\ref{fig:SF}. 
A theoretical expression for the  defocus SF has been derived in KT07.
We  generalize it to  annular apertures  in Appendix~B.   The initial,
quadratic  part of SF is  directly related  to the  combined time
constant $\tau_0$  of all turbulent layers.   However, the acquisition
frequency is not fast enough  to capture the initial quadratic part of
SF  extending only  to time lags  of $<  0.1 t_{\rm cross}  $.  In
order to  get two points  on this  part for a  layer moving with  $V =
50$\,m/s, a frame rate of $\sim 3$\,kHz would be required.

To overcome the sampling problem, we fit the initial part of SF to
a model  of $N$ turbulent  layers with Fried parameters  $r_{0,i}$ and
velocities $V_i$, $1 \leq i\leq N$:
\begin{eqnarray}
D_4(t>0) =  1.94 D^{5/3} 
\sum_{i=1}^{N} r_{0,i}^{-5/3} K_4(2 t V_i/D, \epsilon) 
 +  \frac{2\,  \sigma^2_{\rho, {\rm noise}} }{ A^2},
\label{eq:D4mod}
\end{eqnarray}
where the function $K_4(\beta, \epsilon)$ is defined in Appendix~B and
$\sigma^2_{\rho,  {\rm noise}}$ is  the noise  of the  radius estimate
determined  by  Eq.\,\ref{eq:noise}.    The  adjusted  parameters  are
$r_{0,i}$ and $V_i$.  As will be seen in Sect.\,\ref{Sec:stat1}, the
estimate of $\tau_0$ is  independent of $N$ if $N\geq3$.  Accordingly,
a   three-layer    model   is    chosen   for   the    data   analysis
(Fig.~\ref{fig:SF}).  We  fit only the initial  part of SF, up to
the time increment $\Delta t$.  Its exact value is not critical, as long
as  it is  large enough  for  unambiguous fitting  of the  parameters,
$\Delta t\,\nu  > 2N+1$.   For further data  analysis, we  set $\Delta
t=40$\,ms. 

The   atmospheric  parameters   $(r_0,   \overline{V},  \tau_0)$   are
calculated as
\begin{eqnarray}
r_0^{-5/3} & = & \sum_{i=1}^{N} r_{0,i}^{-5/3} , \label{eq:r0fit} \\
(\overline{V}/r_0)^{5/3} & = & \sum_{i=1}^{N} (V_i/r_{0,i})^{5/3} ,  \\
\tau_0 &=& 0.314 \; r_0/\overline{V} . 
\end{eqnarray}
The estimate of  $r_0$ is also obtained directly  from the ring-radius
variance $\sigma^2_\rho$ by subtracting the noise,
\begin{equation} 
\sigma_\rho^2 - \sigma^2_{\rho,  {\rm noise}}
 = C_\rho (\lambda/D)^2 (D/r_0)^{5/3} .
\label{eq:r0var}
\end{equation}
When SF reaches its  asymptotic value on time  increments smaller
than 40\,ms, the  same value of $r_0$ is  derived from the ring-radius
variance     (Eq.\,\ref{eq:r0var})      and     from     the     model
(Eq.\,\ref{eq:r0fit}).  The robustness  of parameter estimates derived
by  model   fitting  has   been  confirmed  by   numerical  simulation
(Sect.~\ref{Sec:Sim}) and by fitting alternative models to real data
(Sect.\,\ref{Sec:stat1}).

%------------------------------------------------------------------------
\section{Simulations\/}
\label{Sec:Sim}

A new  seeing monitor can be  validated by comparing  it with another,
well-established  instrument. In the case  of FADE,  however, there  is no
reliable  comparison  data  on  $\tau_0$. Instead,  we  simulated  our
instrument  numerically as  faithfully  as we  could  and studied  the
influence of various instrumental and data-reduction parameters on the
final result.

\begin{figure*}[ht]
\begin{center}
\centerline{
\includegraphics[width=8cm]{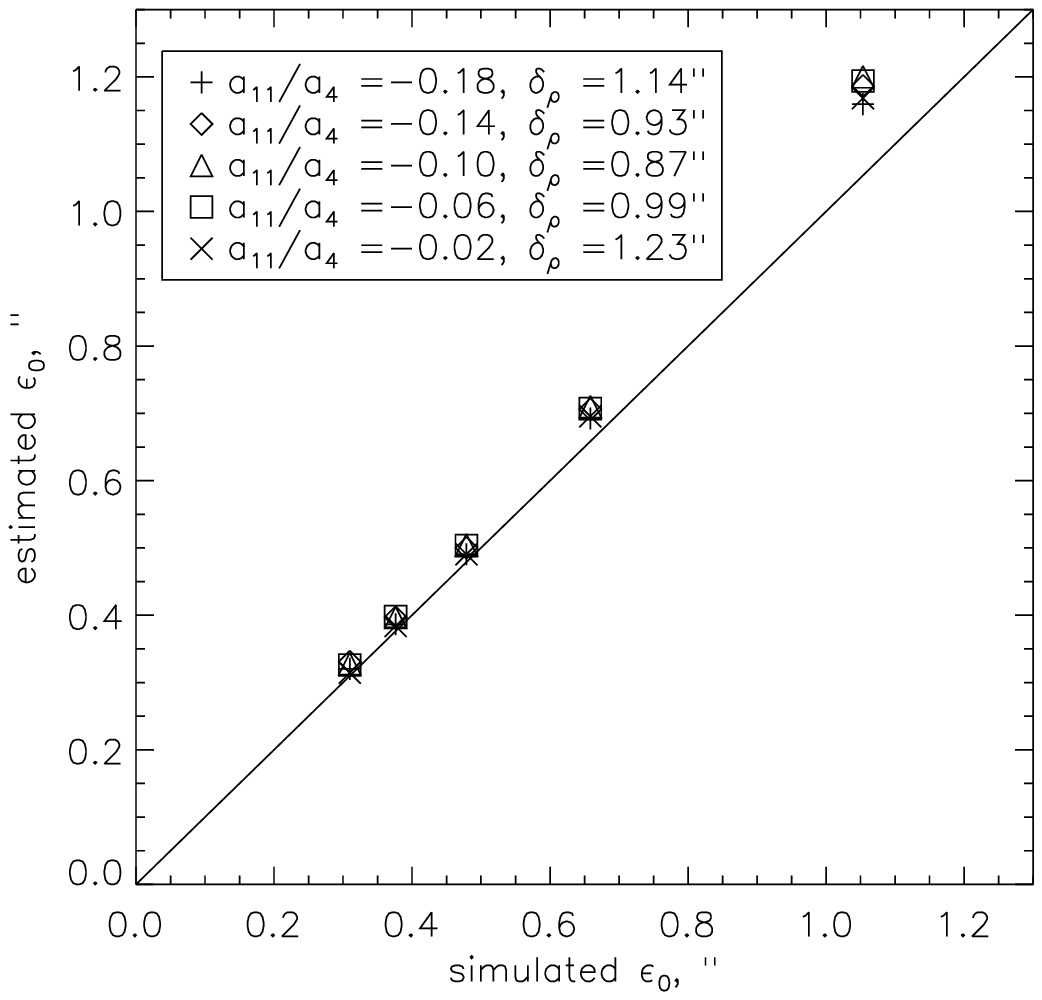} 
\includegraphics[width=8cm]{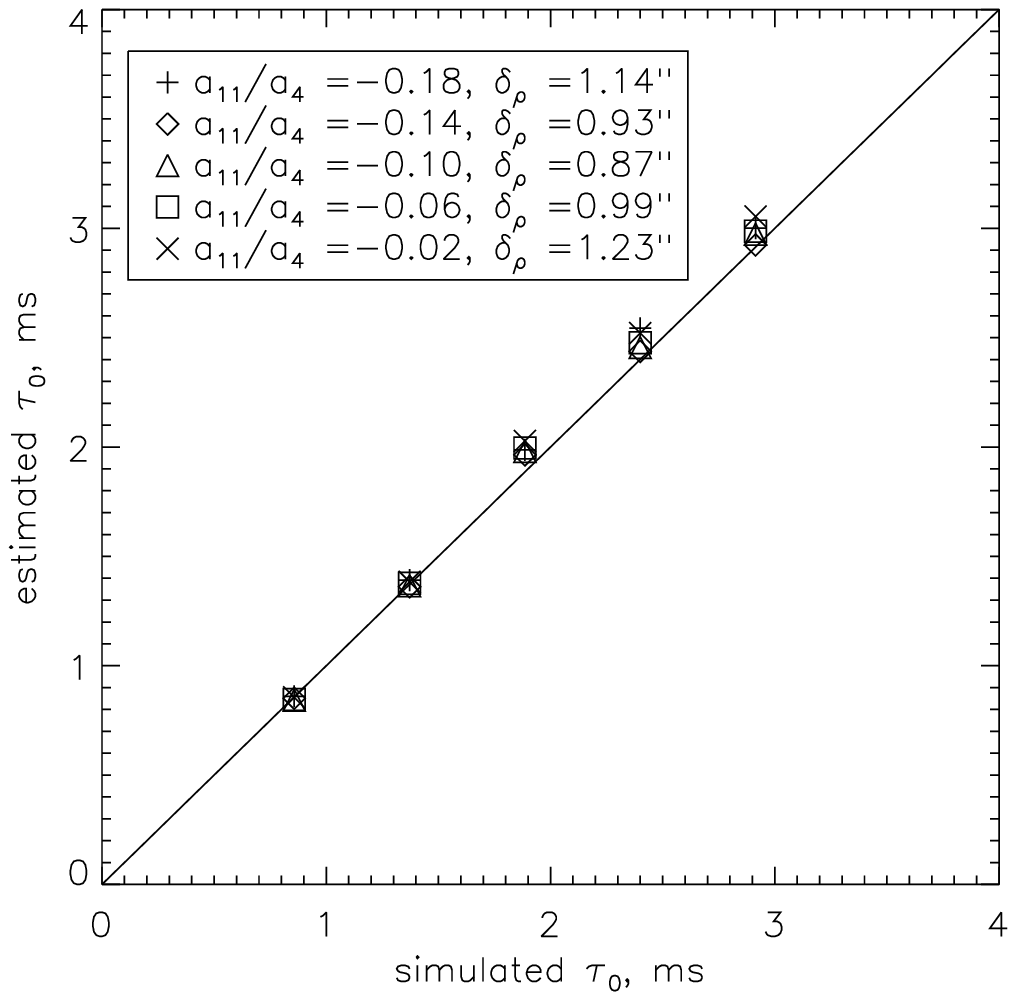} 
}
\centerline{
\includegraphics[width=8cm]{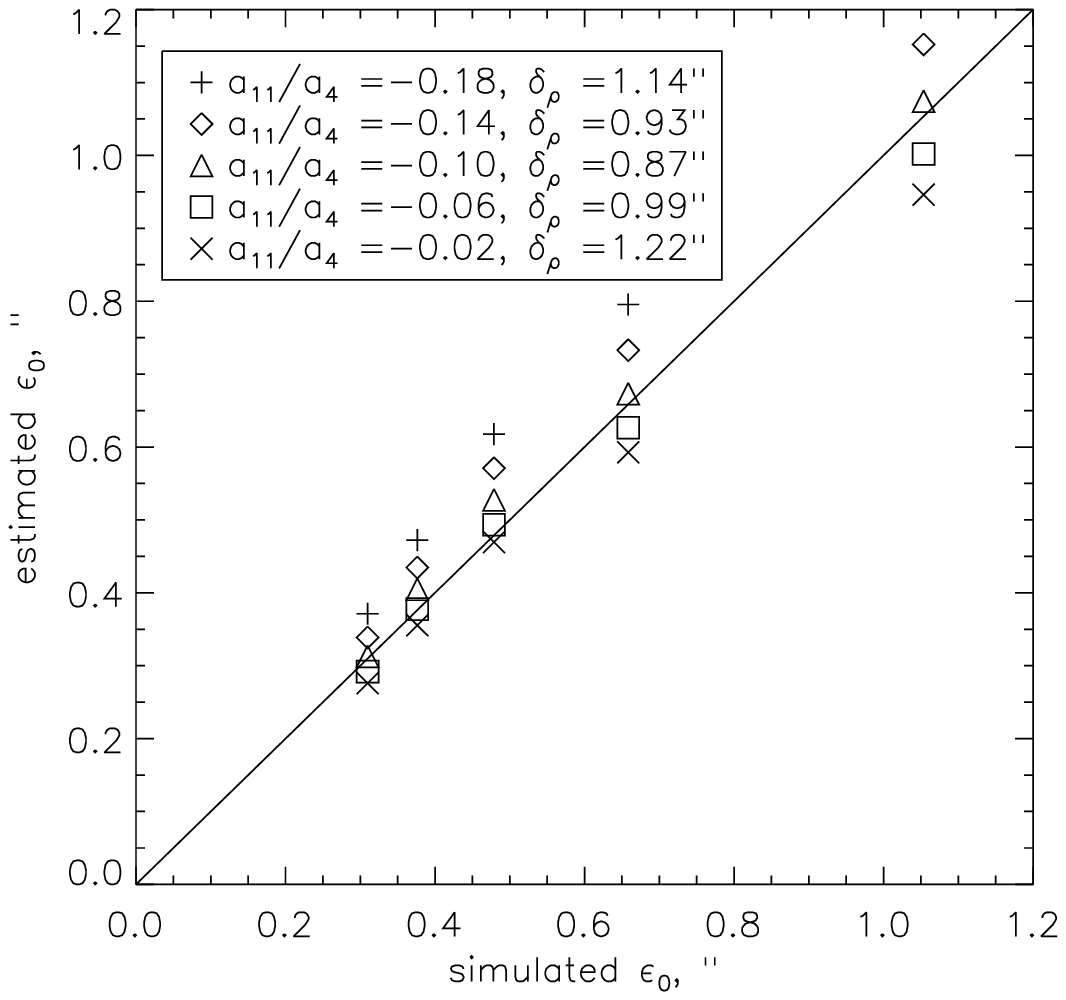} 
\includegraphics[width=8cm]{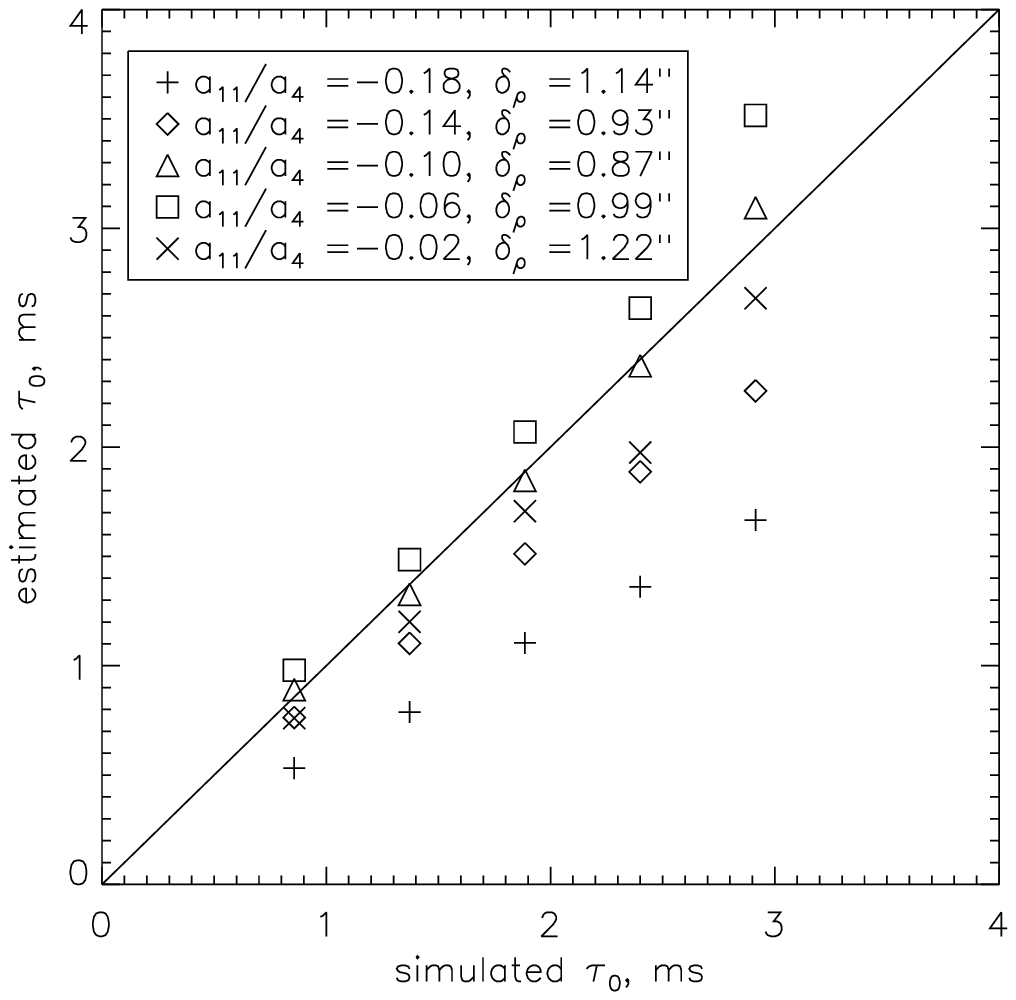} 
}
\caption{ Influence  of the ring  sharpness -- quantified in  terms of
the ring width $\delta_\rho$ --  on the seeing and coherence time estimates.  Top --
turbulence  layer at  ground level,   bottom --  turbulence  layer at
5\,km altitude.  Simulation parameters are given in Table~\ref{tab:simpar}. 
}
\label{fig:spherical} 
\end{center}
\end{figure*}
\begin{figure*}
\begin{center}
\centerline{
\includegraphics[width=5.5cm]{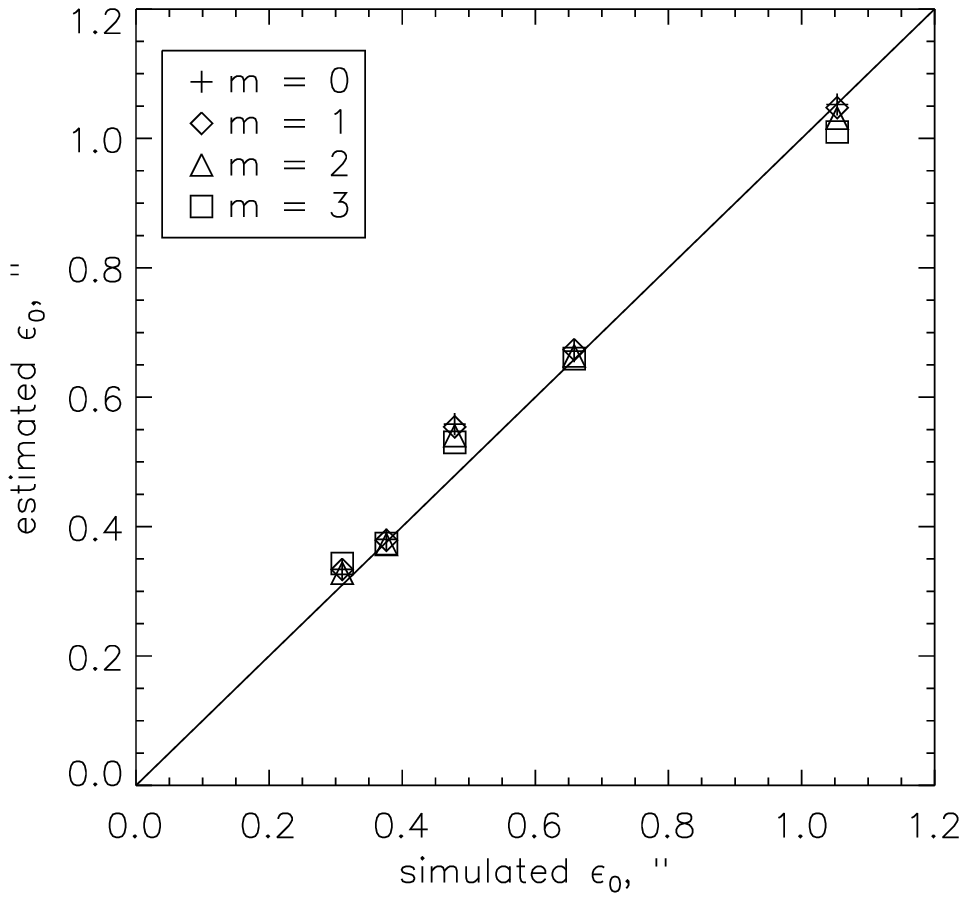} 
\includegraphics[width=5.5cm]{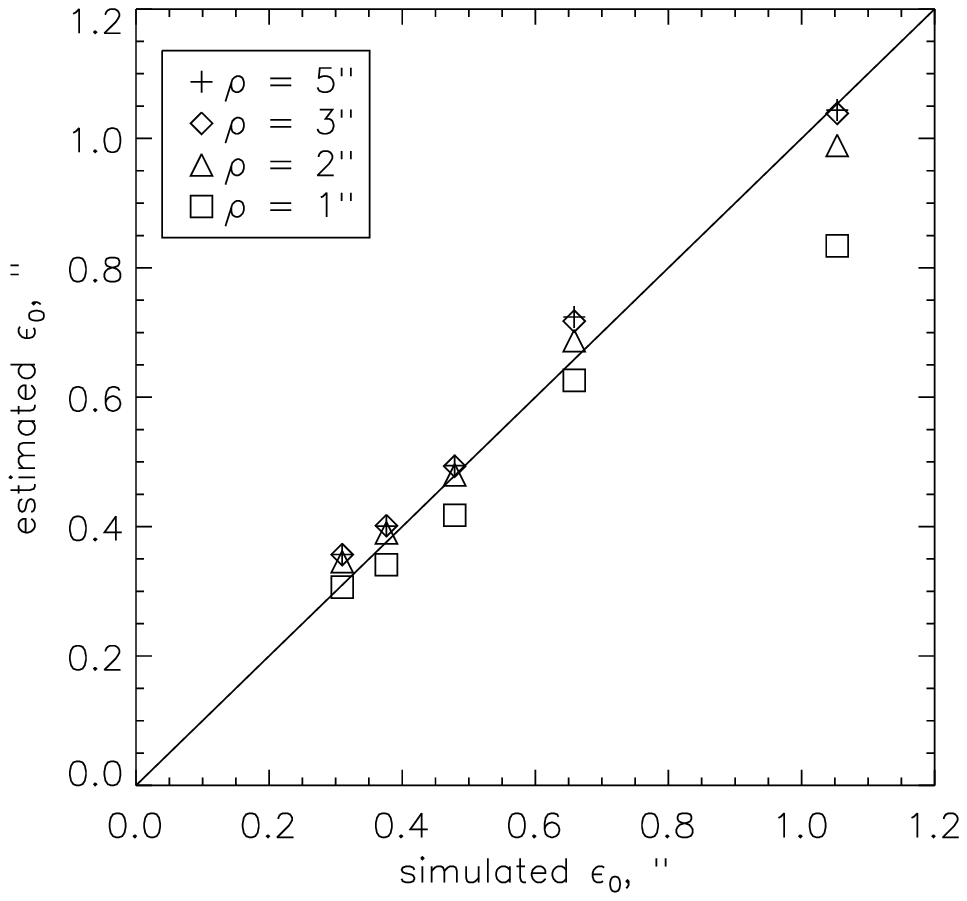} 
\includegraphics[width=5.5cm]{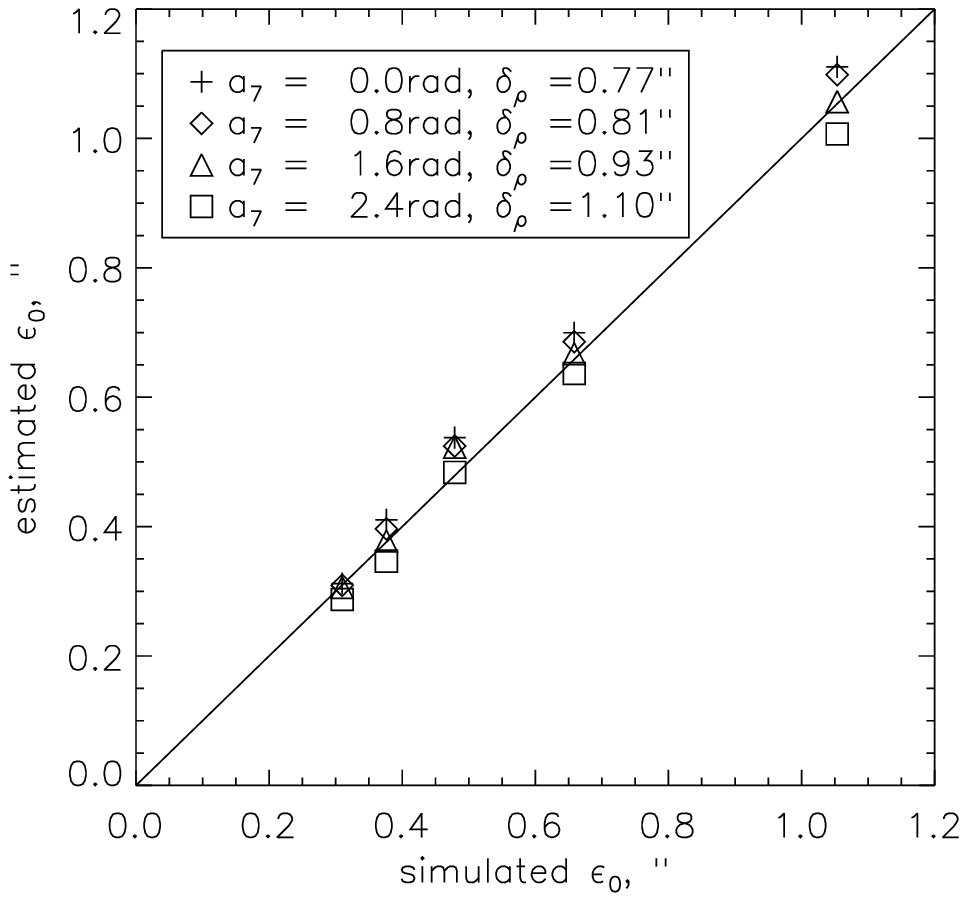} 
}
\centerline{
\includegraphics[width=5.5cm]{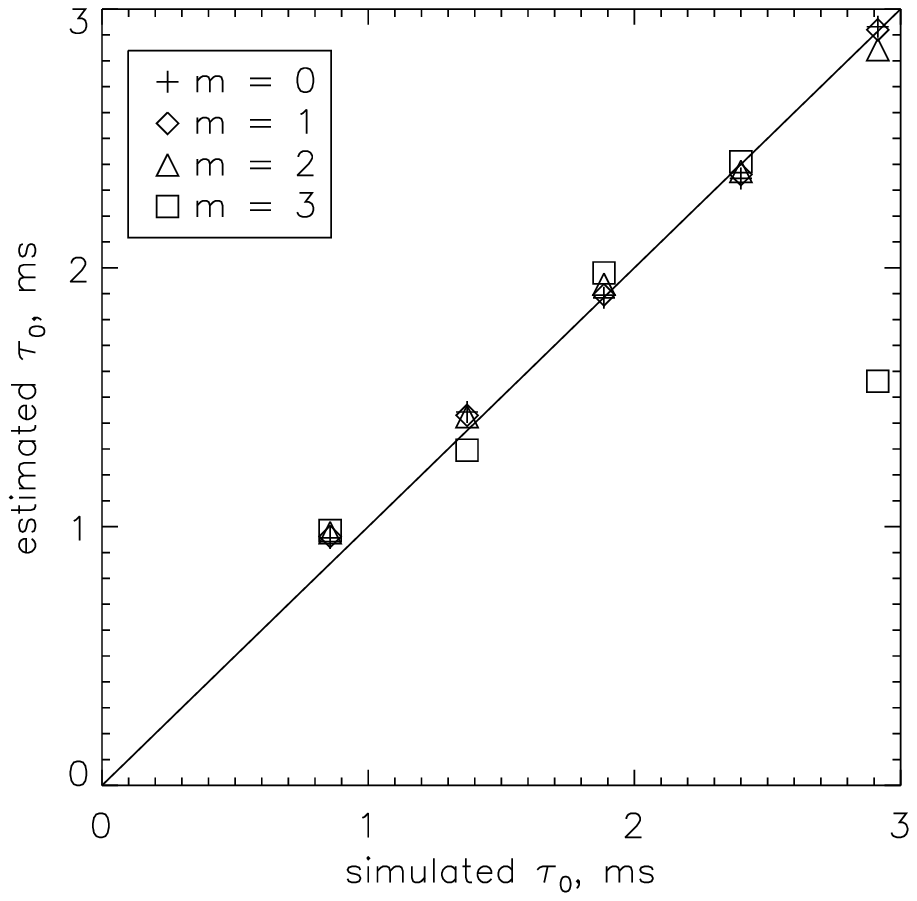} 
\includegraphics[width=5.5cm]{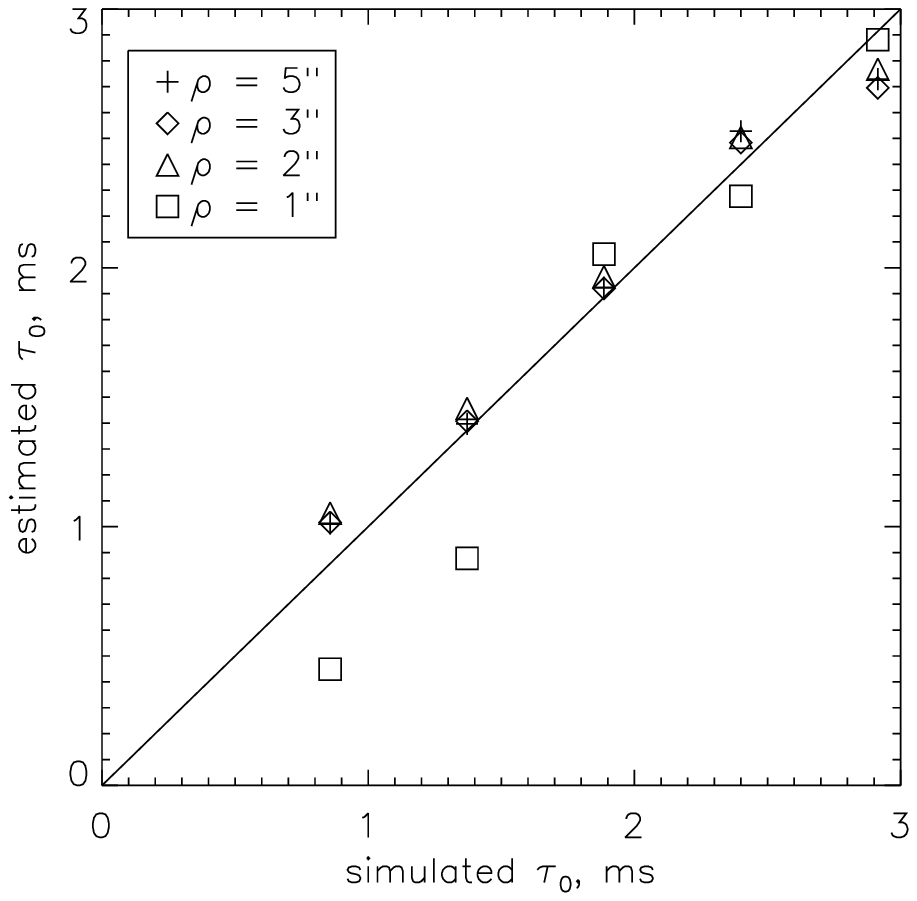} 
\includegraphics[width=5.5cm]{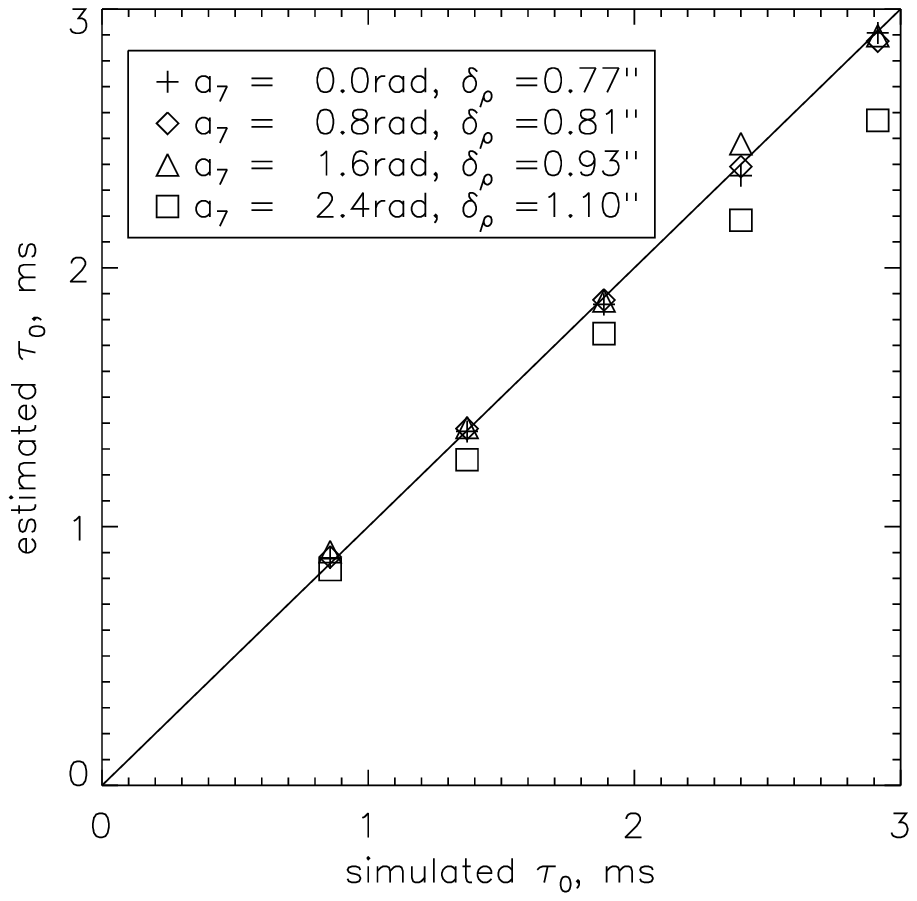} 
}
\caption{Dependence of the seeing and coherence time estimates on stellar magnitude $m_V$,
mean ring radius $\rho$, and coma aberration $a_7$.
Simulation parameters are given in Table~\ref{tab:simpar}. }
\label{fig:uncertainties} 
\end{center}
\end{figure*}

\subsection{Simulation tool}

  Our  simulation tool generates  the complex  amplitude of  the light
field propagated through one  or several phase screens with Kolmogorov
spectrum.   The  screens  are  typically 1024$^2$  pixels  with  1\,cm
sampling, i.e.   about 10\,m across.  The  resulting amplitude pattern
is periodic, without edge effects.   It is ``dragged'' in front of the
simulated telescope with a chosen wind speed, wrapping around edges in
both  coordinates  and  eventually   covering  the  whole  area.   The
monochromatic  images created  by  a telescope  with  a perfect  conic
aberration  of  specified  amplitude  and, possibly,  some  additional
intrinsic  aberrations   are  re-binned  into   the  detector  pixels,
distorted by  readout and  photon noise and  fed to  the data-analysis
routine  instead of  the real  data.  Our  tool has  been  verified by
comparing with analytical results for weak  perturbations and has
been used for simulating other instruments such as DIMM and MASS.  The
limitations  of  this  tool  are the  monochromatic  light,  single  wind
velocity for all layers, and instantaneous exposure time.

We  used two  alternative, nearly  equivalent ways  of  producing ring
images. In  the first method, a perfect conic wavefront  was generated,
and its amplitude was expressed as a ring radius $\rho$. In the second method,
we do not apply conic aberration  ($\rho = 0$), but instead select a
combination  of defocus  and  spherical aberrations  to  mimic a  real
telescope. The sense of the Zernike coefficients $a_4$ and $a_{11}$ in
both cases is distinct.

\subsection{Parameters of the simulations}
\label{Sec:rings}

For convenience, the simulation parameters are gathered in
Table~\ref{tab:simpar}. These parameters are:

\begin{itemize} 
\item[--] number of images in the sequence $N$,
\item[--] acquisition frequency $\nu$,
\item[--] exposure time d$t$,
\item[--] visual stellar magnitude $m$,
\item[--] readout noise $\sigma_{\rm ron\/}$,
\item[--] conic aberration quantified by the average ring radius  $\rho$,
\item[--] amplitudes of the Zernike aberrations $a_4$ (defocus), $a_7$  (coma), $a_{11}$ (spherical), and $a_{27}$, 
\item[--] altitude of the single turbulent layer $h$,
\item[--] seeing $\varepsilon_0$,
\item[--] wind speed $\overline{V}$.
\end{itemize} 

Simulated  ring images  are  compared in  Fig.~\ref{fig:rings} to  the
images of  Sirius recorded on Nov.~2.   The combination of
exposure  time   and  magnitude  results  in  the   detected  flux  of
$3\cdot10^5$\,electrons per  simulated image, as in the  actual images of
Sirius.  For this sequence, the estimated turbulence parameters equal:
$\varepsilon_0=1.05",       \;      \overline{V}      =17$\,m/s,
$\tau_0=1.86$\,ms  (Fig.\,\ref{fig:SF}).    The  same  parameters  are
chosen  for  the  simulated  images.   For  the closest  resemblence  between
simulated  and   real  images,   telescope  aberrations  are   set  to
$a_7=0.7$\,rad,     $a_{11}=-0.75$\,rad,    $a_{27}=0.3$\,rad.     The
turbulence is placed at 13\,km  altitude to reproduce the actual level
of scintillation,  evaluated from the intensity  variance between ring
sectors  $s = 0.011$ (cf. Eq.~\ref{eq:sect}).

\subsection{Refining the response  coefficient}
\label{sec:Acoef}

The coefficient $A$  relating radius variation to defocus  is given by
Eq.~\ref{eq:A}.  Its  actual numerical value, however,  depends on the
method of  radius estimation and, in  particular, on the  choice of the
mask width $\delta$.  For $\delta=4$,  we determined it to equal 1.077
by comparing $\tau_0$ estimates from sequences of simulated images, to
the nominal  input value  of $\tau_0$ when  Eq.~\ref{eq:A} is  used to
relate the radius to defocus.  The corresponding simulation parameters
are summarized in Table\,\ref{tab:simpar}.  Thus,
\begin{eqnarray} 
\Delta \rho /a_4 & = & 1.077 \; \frac{\lambda}{\pi D} \;   
  \frac{\sqrt{12}} {1 - \epsilon}. \label{eq:corr} 
\end{eqnarray}

%-------------------------------------------------------------
\subsection{Instrumental biases}
\label{Sec:uncts}

The data  analysis relies on radius  estimates that can  be altered by
telescope aberrations, scintillation,  detector and photon noise, etc.
Here we  evaluate the instrumental  bias by changing  some parameters,
while other  parameters are  fixed. In each  case, a sequence  of 1024
simulated   images   is    generated   with  the parameter values   listed   in
Table~\ref{tab:simpar}.   The wind  speed is  set to  35\,m/s  and the
coherence time is then changed by modifying the seeing.

\paragraph{Ring sharpness.}
%------------------------------------------
%
The  ring is  sharp  in the  radial  direction when  the wavefront  is
exactly conic. A good approximation of the conic wavefront is achieved
by  the  optimum combination  of  defocus  and spherical  aberrations,
$a_{11}=-0.1\,a_4$.  Here  we explore the  effect of unsharp  ring images by
setting $a_4 =  12$\,rad and varying $a_{11}$ about  its optimum value
$a_{11} =  -1.2$\,rad. Unlike the rest  of the simulations,  we do not
apply conic  aberration and  set $\rho =0$.  Wrong
values of $a_{11}$ make the ring  wider, as shown by its rms width,
$\delta_\rho$ (Eq.~\ref{eq:width}).

As seen  in Fig.\,\ref{fig:spherical}, the seeing and coherence-time
estimates  are  biased in  the case  of blurred rings and  high-altitude
turbulence. The sign of the bias  depends on the sign of the deviation
from  the  optimum  $a_{11}$. When the turbulent layers  are low, the
scintillation  is weak  and the parameters are correctly
derived even if the ring images are blurred.

To ensure  a correct derivation under any  atmospheric conditions, the
ring  width   should  be   close  to  its   diffraction-limited  value
$\delta_{\rho, 0}$:
\begin{eqnarray}
\delta_\rho < 1.2\,\delta_{\rho, 0} \; ; \hskip .2in
\delta_{\rho, 0}=1.7\,\lambda/[D\,(1-\epsilon)] ,
\end{eqnarray}
where  the  coefficient 1.7  is  determined  from  the 
width, $\delta_\rho  =0.87''$, of  diffraction-limited rings.   
Given the instrumental set-up,
images should  be rejected  if $\delta_\rho<1''$. However,  all images
recorded with the FADE prototype have  $\delta_\rho > 1''$. Hence, we apply a
softer data-selection criterion: $\delta_\rho  < 1.25''$, and note that the
resulting estimates might still be biased if the turbulence was high.

\paragraph{Stellar magnitude, ring radius, and coma.}
%--------------------------------------
%
Figure\,\ref{fig:uncertainties} examines  the stability of  the seeing
and coherence  time estimates with  respect to the  stellar magnitude
$m_V$,  ring  image radius  $\rho$,  and  coma  aberration $a_7$.   In
agreement  with  Eq.\,\ref{eq:noise},  estimates  are  correct  up  to
stellar magnitudes  2--3.  Spatial sampling and coma  aberration do not
affect the estimates  if $\rho \geq 2''$ (i.e. 5  pixels) and $a_7 \leq
2$\,rad.

%------------------------------------------------------------
\section{Analysis of observations}\label{statistics} 

Seeing and coherence time   were estimated from all sequences
of   4000  images  recorded   with  FADE   at  Cerro   Tololo  between
October\,29  and November\,2,  2006.  In  this
section, we  check the FADE  results for consistency and  compare them
with the MASS-DIMM.

\subsection{Influence of instrumental parameters}
\label{Sec:stat1}

During data  acquisition, instrumental  parameters were varied  over a
broad range to evaluate their  effect on the results.  Even though the
non-stationarity of the  atmosphere precludes direct comparisons, some
conclusions can nevertheless be drawn.

The {\it ring sharpness}, $\delta_\rho$,  has been identified as a major
source  of instrumental  bias in  FADE when  significant high-altitude
turbulence  is   present.   In  our   data,  most  images   have  $1''
<\delta_\rho < 1.5''$, whereas  a perfect diffraction-limited ring has
$\delta_\rho= 0.9''$.   Analysis of  the average ring images confirms
that the optimum combination  of defocus and spherical aberrations was
not reached,  $a_4$ and $a_{11}$ often  having  the  same sign rather
than   opposite signs.   For our  data, the  dispersion and  mean of
$\tau_0$  increase  when  $\delta>1.25''$; accordingly,  sequences  with
$\delta_\rho>  1.25''$ are  disregarded.   Still, some  bias caused  by
radially defocused images remains.
\begin{figure}[ht]
\begin{center}
\includegraphics[width=8.5cm]{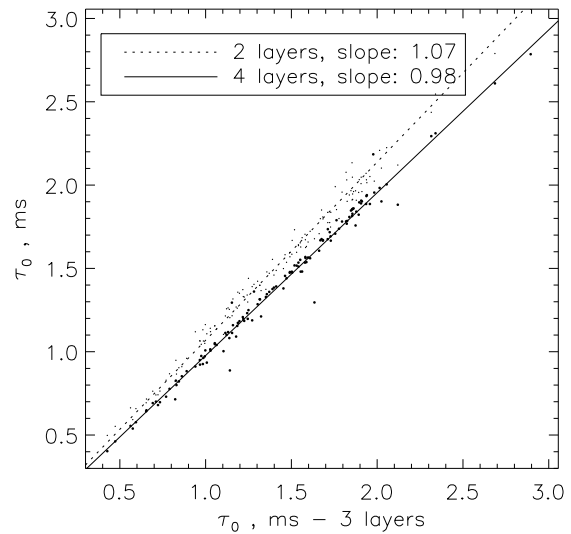} 
\caption{ Coherence time  derived from the recorded  data by fitting
a model with $N= 3$  layers (x-axis)  compared to the coherence
time derived with 
$N=2$ (dotted line) and  $N=4$ (solid line) models.  }
\label{fig:nlayers}
\end{center}
\end{figure}

As  described in Sect.\,\ref{sec:model},  the data  are fitted  to a
model with a discrete {\it  number of turbulent layers\/}, $N$.  What 
is  the  minimum value  of  $N$  that  permits a correct  derivation  of
$\tau_0$?   Figure\,\ref{fig:nlayers} shows  that the  $\tau_0$ values
obtained with 3 and 2 (resp. 4) layers differ on average by 7\%
(resp. 2\%).  An  average difference of 2\% is  likewise obtained when
comparing the estimates  with  3 and 5  layers.  A 3-layer model
is  thus a  good compromise  enabling a  fit to the  data with  only six
parameters.
\begin{figure}[ht]
\begin{center}
\includegraphics[width=8.5cm]{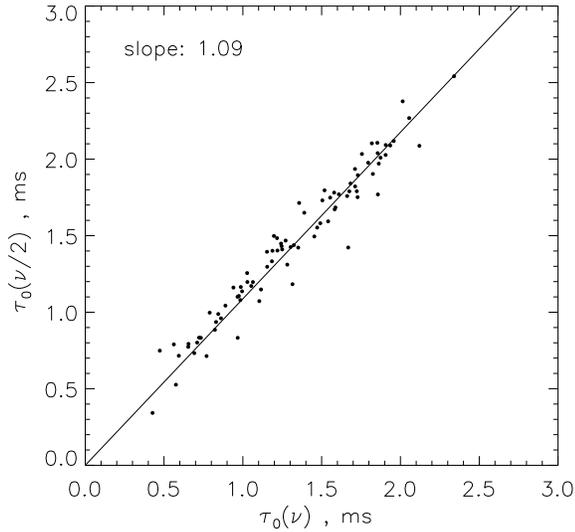} 
\caption{  Coherence time  derived  from all sequences  recorded with 
$\nu \ge 700$\,Hz  when every image  (x axis) and  every other  image (y
axis) is considered. 
\label{fig:interleave} }
\end{center}
\end{figure}

The  influence of  the  {\it acquisition  frequency}  on the  measured
coherence  time is examined  in Fig.\,\ref{fig:interleave}.   The data
sequences recorded  at frequencies $\nu \ge  700$\,Hz were re-analyzed
considering  every other image.   The coherence  time obtained  with a
slower $\nu/2$  sampling is  on  average  9\% longer than  with the
fast  sampling. This difference  is reproduced  by simulations  if the
turbulence  is placed  at 5\,km  altitude and  if the  ring-images are
slightly defocused  in the radial  direction ($a_{11}/a_4\approx-0.07$
or  $-0.14$ instead  of  $a_{11}/a_4=-0.11$ corresponding  to a  sharp
ring).  The  effect of temporal  under-sampling is perceptible  if the
same  comparison is  repeated with  sequences recorded  at frequencies
below 700\,Hz: the number of points on the initial, increasing part of
SF is then not  always sufficient to unambiguously extract the six
fitted parameters, and the  coherence time is  poorly constrained.
To  ensure  a correct  temporal  sampling  under  fast turbulence,  we
ignored the sequences with $\nu < 500$\,Hz.

In  line with  the simulations,  the coherence  time estimates  do not
depend  on  the  average  {\it  ring image  radius}.   Similarly,  the
parameter statistics seem  unbiased by the {\it stellar  flux} and by
the {\it exposure  time}. While the sequences of  Sirius ($m_V =-1.5$)
and Fomalhaut ($m_V =1.2$) images were recorded with exposure times of
d$t<0.5$\,ms  and  $1.0<{\rm d}t  <1.9$\,ms,  respectively, no  obvious
difference exists between the mean  and rms of the atmospheric
parameters measured in terms of these  two stars, \\
$\epsilon_0^{\rm S}:\;(0.9\pm0.2)'' \hskip .4in
\epsilon_0^{\rm F}:\;(0.8\pm0.1)''$ \\ 
$\tau_0^{\rm S}:\;(1.4\pm0.5)$\,ms \hskip .25in
$\tau_0^{\rm F}:\;(1.3\pm0.5)$\,ms.\\ 

\subsection{Comparison with MASS and DIMM}

In this section, the seeing  and coherence time obtained with FADE are
compared to simultaneous measurements by the CTIO site monitor located
at  10\,m  distance from  FADE  on a  6\,m  high  tower.  The  monitor
consists of  a combined  MASS-DIMM instrument fed  by the  25-cm Meade
telescope and looking at bright ($V=2^m...3^m$) stars near zenith.  Of
particular interest  here is the  time constant $\tau_0$  estimated by
MASS from the temporal  characteristics of scintillation by the method
of Tokovinin  (\cite{MASS_tau}).  This method  is intrinsically biased
because it  does not account  for the turbulence below  $\sim 500$\,m.
Moreover,  it has been  recently established  by simulations  that the
coefficient used  to calculate $\tau_0$  in the MASS software  must be
increased  by 1.27.\footnote{See the  unpublished report  by Tokovinin
(2006) at                                                            \\
http://www.ctio.noao.edu/\~{}atokovin/profiler/timeconst.pdf}  In  the
following,  we  correct  $\tau_0$  by applying  this  coefficient  and
including the contribution of the ground layer:
 \begin{eqnarray}
 \tau_0^{-5/3} &=& (1.27\;\tau_{\rm MASS\/})^{-5/3} + 
118\,\lambda^{-2}\,V^{5/3}_{\rm GL\/}\,(C_{\rm n}^2\,{\rm d}h)_{\rm
  GL} .
\label{eq:MASScor}
 \end{eqnarray}
The  turbulence integral  in the  ground layer,  $(C_{\rm n}^2\,{\rm
d}h)_{\rm  GL\/}$,  is  computed   from  the  difference  between  the
turbulence  integrals measured  by  DIMM (whole  atmosphere) and  MASS
(above 500\,m), while the ground  layer wind speed, $V_{\rm GL\/}$, is
known from the local meteorological station.  Even after correction by
Eq.\,(\ref{eq:MASScor}),  the  coherence  time measured  by  MASS-DIMM
should  be taken  with  some  reservation because  it  has never  been
checked against independent instruments and some bias is possible.

\begin{figure}[ht]
\begin{center}
\includegraphics[width=8.5cm]{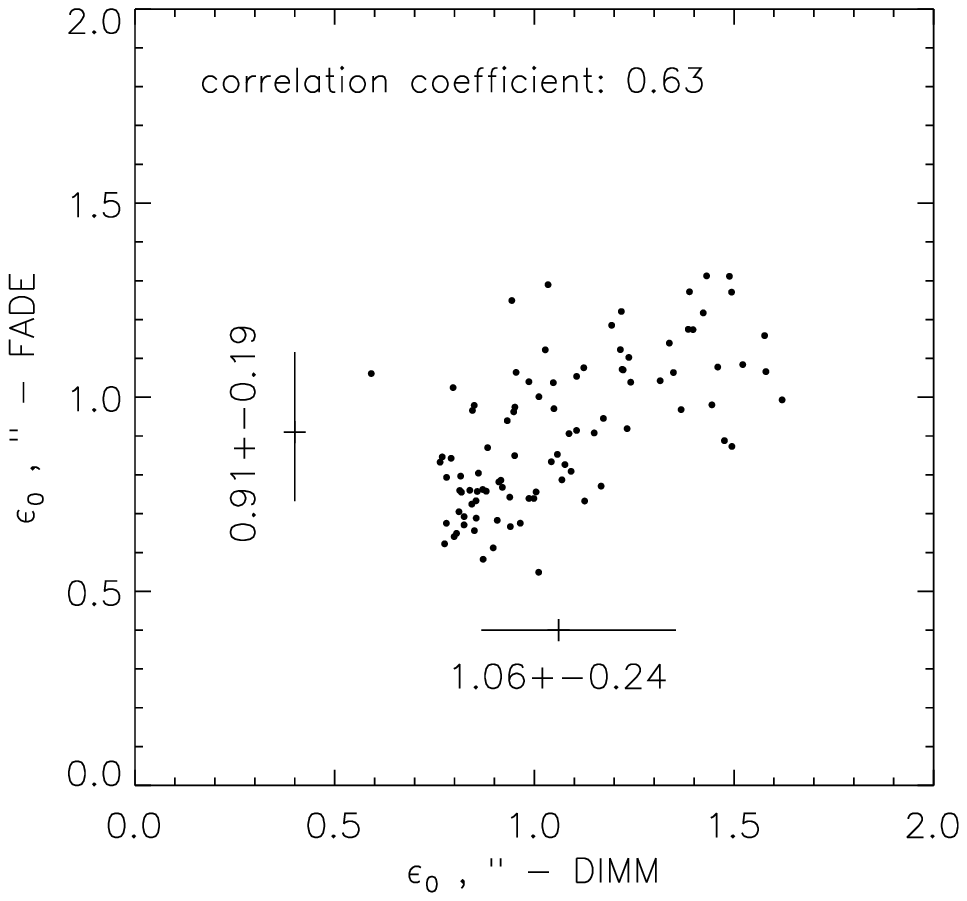} 
\includegraphics[width=8.5cm]{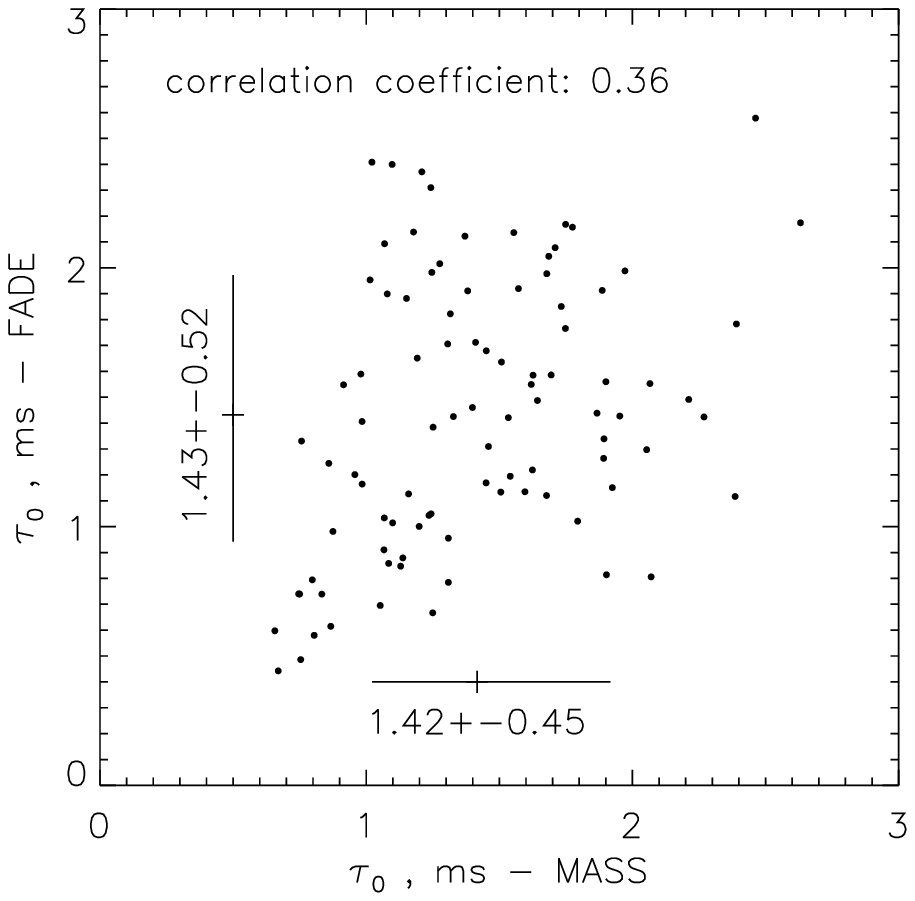} 
\caption{Seeing and coherence time  measured with FADE between October
29 and  November  2,  compared to  simultaneous
measurements  by  the  MASS-DIMM.   The average  values  and  standard
deviations  of  parameters   and  the  correlation  coefficients  are
indicated.  }
\label{fig:eps_tau} 
\end{center}
\end{figure}

Figure\,\ref{fig:eps_tau}  compares the  estimates  of $\varepsilon_0$
and $\tau_0$ obtained with FADE from October\,29 to November\,2 
to the results  of MASS-DIMM.  We do not expect detailed
correlation,   because   the   instruments  were   sampling   different
atmospheric volumes.  As seen in Fig.\,\ref{fig:eps_tau}, the 
seeing measurements are  better  correlated  than the coherence  times.

Statistically,  it  appears  that  FADE  slightly  underestimates  the
seeing.  This  effect is reproduced with  simulations of high-altitude
turbulence   if  the   ratio  of   spherical  aberration   to  defocus
$a_{11}/a_4$ is set higher than its optimum value $-0.1$ corresponding
to  sharp ring  images.  In  this case,  FADE also  underestimates the
coherence time.  The bias on $\tau_0$ can, however, not be ascertained
by  Fig.\,\ref{fig:eps_tau}  because the  $\tau_0$  estimates by  MASS
might  likewise be  biased.  Thus,  the comparison  presented  in this
section cannot be considered as a validation of FADE.

\section{Conclusions and perspectives}
\label{Conclusions}

We have  built a  first prototype of  the site-testing  monitor, FADE,
suitable for  routine measurements  of the atmospheric  coherence time
$\tau_0$, as  well as the  seeing $\varepsilon_0$. The  instrument was
  tested  on  the  sky.  Extensive  simulations  substantiate  the
validity  of  the FADE  results  and  indicate potential  instrumental
biases.  Our main conclusions are as follows:

\begin{itemize}
\item
The sampling  time of the image  sequence must be a  small fraction of
the aperture crossing  time $t_{\rm cross} = D/V$  ($ \sim$10\, ms for
$D=0.36$\,m and wind speed $V=36$\,m/s). Sampling at $\nu \ge 500$\,Hz
appears adequate under most conditions.

\item
The sharpness of the ring image  in the radial direction does not bias
the results when the turbulence is located near the ground. But it can
bias  both $\tau_0$  and  $\varepsilon_0$ estimates  when high  layers
dominate, and {\it strict control of the telescope aberrations is thus
required.}  The aberrations (hence the data validity) can be evaluated
{\it a  posteriori} from the average ring  image.  Real-time estimates
of the  ring radius  $\rho$ and width  $\delta_\rho$ are  needed to
ensure good optical adjustment of the instrument.

\item
The FADE  monitor with 36-cm telescope  can work on stars  as faint as
$m_V =3^m$.

\item
A simple estimator of  the ring radius (Eq.\,\ref{eq:rho}) is adequate
and robust, provided a wide  enough mask around the ring ($\delta \sim
4$) is used in the calculation.

\item
Moderate telescope aberrations such as coma are acceptable. The
results are not critically influenced by small telescope focus errors.  

\end{itemize}

The current  FADE prototype stores  all image sequences, leading  to a
large  data  volume; the  data  are  processed  offline.  While  this
procedure was necessary for  the first experiments, online processing
will be  implemented in a  definitive instrument.  We  have formulated
and tested the data processing  algorithm and can now develop adequate
real-time software.

We plan to develop an improved version of FADE with real-time
data  analysis.  It will  be  compared  to  simultaneous estimates  of
the atmospheric time constant from currently working adaptive-optics systems
(Fusco  et al.   \cite{Fusco2004})  and/or long-baseline  interferometers
such  as  VLTI.   Characterization   of  Antarctic  sites  for  future
interferometers is an obvious application for FADE.

\begin{acknowledgements}

This work  was stimulated by  discussions with 
Marc    Sarazin    and    other    colleagues   involved    in    site
characterization. We acknowledge financial  and logistic help from ESO
in  building and  testing the  first FADE  prototype.  We  thank the Cerro
Tololo Inter-American  Observatory for its hospitality  and support of
the first FADE mission.

\end{acknowledgements}

%-------------------------------------------------------------------
\appendix

%-------------------------------------------------------------------
\section{Estimator of the ring radius and center}
\label{sec:radius}

The parameters  of the ring-like  image -- its center  $(x_c,y_c)$ and
radius $\rho$ --  can be derived by minimizing  the intensity-weighted
mean squared distance of the pixels from the circle, $\delta^2_{\rho} $:
\begin{eqnarray}
	\delta^2_{\rho} &=& \sum_{l,k}  I_{l,k}\;(r_{l,k} - \rho )^2 \; / \; \sum_{l,k}  I_{l,k},	        
\label{eq:sigma2}
\end{eqnarray}
where $r_{l,k} = [ (l - x_c)^2 + (k - y_c)^2]^{0.5}$ is the distance of pixel
$(l, k)$ from the ring center, $(x_c,y_c)$.
Setting the  partial derivative of $\delta^2_{\rho}$ over  $\rho$ to zero,
we  obtain  the radius  estimator of Eq.\,\ref{eq:rho}.  However, it  still
depends  on the unknown  parameters $(x_c,y_c)$.  
By use of Eq.\,\ref{eq:rho}, Eq.\,\ref{eq:sigma2} is simplified to:
\begin{eqnarray}
\delta^2_{\rho} & = &
\frac{\sum_{l,k} I_{l,k} r_{l,k}^2}{\sum_{l,k} I_{l,k} }   -  \;  
\left [ \frac{\sum_{l,k} I_{l,k} \; r_{l,k}}{\sum_{l,k} I_{l,k}} \right]^2 .  
\label{eq:sig2a}
\end{eqnarray}
This  formula  does  not   contain  $\rho$.   The  center  coordinates
$(x_c,y_c)$  can be  derived  by setting  the  partial derivatives  of
$\delta^2_{\rho} $  over parameters  to zero  and solving  the  equations.  We
determine  the center numerically  by minimizing  Eq.\,\ref{eq:sig2a} and
using  the center-of-gravity coordinates as a starting point.

%-------------------------------------------------------------------
\section{Structure function of atmospheric defocus}
\label{sec:SF}

The  temporal structure  function  of atmospherically-induced  defocus
variations  --  Zernike  coefficient  $a_4$  in  Noll's  (\cite{Noll})
notation -- has  been derived in KT07 for  a filled circular aperture.
Here we generalize  it to an annular aperture.   Without repeating the
whole  derivation, we refer  the reader  to KT07  and modify  only the
spatial  spectrum   of  the   Zernike  defocus,  taking   the  central
obstruction ratio $\epsilon$ into account. The resulting expression is
\begin{eqnarray}
D_{4} (t) & = & 0.821 \;  k^2 D^{5/3}
\int_0^{+\infty}  {\rm d\/}h \;  C_{\rm n}(h)^2  
\; K_4 \left( \frac{2 t V(h)}{ D}, \epsilon \right) ,\\ \label{eq:D4}
K_4(\beta, \epsilon) & = & \frac{12}{\;(1-\epsilon^2)^4}\; 
\int_0^{+\infty}  {\rm d\/}x \; x^{- 8/3} \;[1-J_0(\beta x)]\;  \nonumber \\
& \times &  \left[ 
\frac{J_3(x)}{x} - \epsilon^4 \frac{J_3(\epsilon x)}{\epsilon x} + \;
\epsilon^2  \frac{J_1(x)}{x} - \epsilon^2  \frac{J_1(\epsilon
  x)}{\epsilon x} 
\right]^2 ,
\label{eq:K4}
\end{eqnarray}
where $k=2\pi/\lambda$, $J_n$ is the Bessel function of  order $n$,
$C_{\rm n}(h)^2$   and   $V(h)$   are   the   altitude   profiles   of   the
refractive-index  structure constant  and  wind speed,  respectively.
Considering the known relation between the turbulence integral and the
Fried parameter, $r_0^{-5/3} = 0.423\,k^2\,C_{\rm n}^2 {\rm d}h$, we can also
write the  defocus SF produced by a {\em single layer} as  
\begin{eqnarray}
D_{4} (t) = 1.94 \, (D/r_0)^{5/3}  K_4 ( 2 t V/ D, \epsilon ) .
\label{eq:D4a}
\end{eqnarray}
For calculating  the function $K_4$,  it is convenient  to approximate
the integral (\ref{eq:K4})  by an analytical formula, as  in KT07.  We
suggest the approximation
\begin{eqnarray}
K_4 (\beta, \epsilon) \approx \frac{C_1\, \beta^2 + C_2 \,\beta^6} 
{1 + C_3 \,\beta^\alpha + \beta^6} ,
\label{eq:K4apr}
\end{eqnarray}
where  the coefficients are  cubic polynomials  of $\epsilon$:
\begin{eqnarray}
C_i = C_{i,0}          \sum_{k=0}^{3}          c_{i,k}     \,     \epsilon^k,
\end{eqnarray}
cf. Table~\ref{tab:coef}. This approximation  is valid for $\epsilon <
0.6$ with a maximum relative error of less than 5\% (3\% for $\epsilon
=0.42$)  and correct  asymptotes.  The  asymptotic  value $K_4(\infty,
\epsilon)  =  C_2$  gives  the  focus variance  on  annular  aperture,
analogous to the Noll's coefficient. For $\epsilon=0$, we  get $C_2 =
0.024$  and the focus  variance coefficient  of $1.94\times  0.024/2 =
0.0233$, in agreement with Noll's result. 

\begin{table}[ht]
\center
\caption{Coefficients of (\ref{eq:K4apr}) }
\label{tab:coef}
\medskip
\begin{tabular}{l  c c c c c }
\hline
\hline
Param. & $C_0$ & $\epsilon^0$ & $\epsilon^1$ & $\epsilon^2$ & $\epsilon^3$ \\
\hline
$C_1$ & 0.04642 & 1 & $-$0.182 & $-$2.431 & 2.028 \\
$C_2$ & 0.0240  & 1 & $-$0.017 & $-$3.619 & 2.833 \\
$C_3$ & 1       & 1.25 & 0     &   0      & 7.5 \\
$\alpha$ &   1  & 2.18 & $-0.93$&  0      &  0 \\ 
\hline
\hline
\end{tabular}
\end{table}

The function $K_4 (\beta, \epsilon)$ reaches half its saturation value
at $\beta = 0.63$; hence, the atmospheric defocus
correlation time is $\sim 0.3 D/V$, as is well known in adaptive
optics. 

The above analysis is  valid for instantaneous measurements, while the
defocus is  in fact averaged over  the exposure time.   This effect is
usually non-negligible for the DIMM.  The time averaging can be included as
an additional  factor in the  integral (\ref{eq:K4}), as done  e.g. in
Tokovinin (\cite{MASS_tau}).  We made  this calculation and found that
the  initial,  quadratic  part  of  $D_4(t)$  (or,  equivalently,  the
parameter  $C_1$  in Eq.~\ref{eq:K4apr})  is  reduced  by  0.8 for  an
exposure time $t_{\rm  exp} \sim 0.3 D/V$ and a  layer moving with the
speed $V$. To adequately sample the  SF features produced by
the  fastest-moving layers,  the sampling  time (hence  exposure time)
must be  shorter than $ 0.3  D/V_{\rm max}$, so the  bias caused by
the finite exposure in FADE  can be neglected.  In  hindsight, this
result could be expected:  to follow the focus variations, we
need such  a fast  sampling that the  integration during  the sampling
period has a negligible effect.

%-------------------------------------------------------------------

{}


\begin{thebibliography}{}

\bibitem[2004]{Fusco2004}
Fusco, T., Ageorges, N., Rousset, G., Raboud, D., 
Gendron, E., Mouillet, D., Lacombe, F. et al.
%Zins, G., Charton, J., Lidman, C., Hubin, N.
 2004, Proc. SPIE, 5490, 118


\bibitem[2007]{Fade}
Kellerer, A., \& Tokovinin, A. 
2007, A\&A,  461, 775 (KT07)


\bibitem[2003]{MASS1}
Kornilov, V., Tokovinin, A., Vozyakova, O., 
Zaitsev, A., Shatsky, N., Potanin, S., \& Sarazin, M. 
2003, 
Proc. SPIE, 4839, 837


\bibitem[1992]{Lopez92}
Lopez, B.  
1992,  A\&A, 253, 635


\bibitem[1976]{Noll}
Noll, R. 1976, J. Opt. Soc. Am., 66, 207

\bibitem[2003]{Perrin}
Perrin, M.D., Sivaramakrishnan, A., Makidon, R.B. et al.  2003, ApJ, 596, 702

\bibitem[1981]{Roddier81}
Roddier, F.
1981, 
Progress in Optics, 19, 281


\bibitem[1990]{DIMM}
Sarazin, M., \& Roddier, F. 
1990,  A\&A, 227, 294


\bibitem[2002]{MASS_tau}
Tokovinin, A. 2002,  
Appl. Opt.,  41, 957


\bibitem[2006]{Donut}
Tokovinin, A. \& Heathcote, S. 2006, 
PASP,  118,  1165

\end{thebibliography}
\end{document}